%% file: main_arXiv.tex
\documentclass[pra,aps,twocolumn,amsmath,footinbib]{revtex4-2}

\usepackage{amssymb} 
\usepackage{mathrsfs}
\usepackage{graphicx} 
\usepackage{color}

\makeatletter
\def\maketitle{
\@author@finish
\title@column\titleblock@produce
\suppressfloats[t]}
\makeatother

\usepackage[caption=false]{subfig}
\newcommand{\phantomsubfloat}[1]{
    {
        \captionsetup[subfigure]{labelformat=empty}
        \subfloat[][]{#1}
    }%
}

\begin{document}   

\title{Bridging Effective Field Theories and Generalized Hydrodynamics}

\author{F. M{\o}ller$^1$} 
\author{S. Erne$^1$}
\author{N. J. Mauser$^{2}$} 
\author{J. Schmiedmayer$^1$}
\author{I. E. Mazets$^{1,2,3}$} 
\affiliation{$^1$ Vienna Center for Quantum Science and Technology (VCQ), Atominstitut, TU Wien, 1020 Vienna, Austria}
\affiliation{$^2$ Research Platform MMM ``Mathematics---Magnetism---Materials," \\ c/o Fakult\"at f\"ur Mathematik,
Universit\"at Wien, 1090 Vienna, Austria} 
\affiliation{$^3$ Wolfgang Pauli Institut,  c/o Fakult\"at f\"ur Mathematik,
Universit\"at Wien, 1090 Vienna, Austria}

\begin{abstract} 
Generalized Hydrodynamics (GHD) has recently been devised as a method to solve the dynamics of integrable quantum many-body systems beyond the mean-field approximation. In its original form, a major limitation is the inability to predict equal-time correlations. Here we present a new method to treat thermal fluctuations of a 1D bosonic degenerate gas within the GHD framework. We show how the standard results using the thermodynmaic Bethe ansatz can be obtained through sampling of collective bosonic excitations, revealing the connection or duality between GHD and effective field theories such as the standard hydrodynamic equations. As an example, we study the damping of a coherently excited density wave and show how equal-time phase correlation functions can be extracted from the GHD evolution. Our results present a conceptually new way of treating fluctuations beyond the linearized regime of GHD.

\end{abstract} 

\maketitle

\input{manuscript}

This project was supported by the John Templeton Foundation under Grant ID 62179  'Bridging physical theories' and by the DFG/FWF Collaborative Research Centre `SFB 1225 (ISOQUANT)'.
N.M. and I.M. acknowledge the support by the Wiener Wissenschafts- und Technologiefonds (WWTF) via project No. MA16-066 (SEQUEX) and the Austrian Science Foundation (FWF) via project No. F65 (SFB “Taming complexity in PDE Systems”).
S.E. acknowledges support through an ESQ (Erwin Schrödinger Center for Quantum Science and Technology) fellowship funded through the European Union’s Horizon 2020 research and innovation program under Marie Skłodowska-Curie Grant Agreement No 801110. This project reflects only the authors’ view, the EU Agency is not responsible for any use that may be made of the information it contains. ESQ has received funding from the Austrian Federal Ministry of Education, Science and Research (BMBWF). 


\appendix



\input{supplemental_material}

\bibliography{references}

\end{document}

%% file: manuscript.tex

Ultracold atomic gases confined to an effectively one-dimensional (1D) geometry~\cite{PhysRevLett.105.265302, PhysRevLett.87.130402, doi:10.1126/science.1100700} have become an important tool to study the out-of-equilibrium dynamics of many-body quantum systems~\cite{RevModPhys.80.885}. 
The absence of Bose-Einstein condensation and long-range order in 1D systems~\cite{DalfovoRev, Popov87} stems from fluctuations playing an important role, even at low temperatures.
In experiments, fluctuations in the density have been used for characterization~\cite{PhysRevLett.96.130403, PhysRevLett.105.230402, PhysRevLett.106.230405}, while phase fluctuations give rise to the formation of density ripples in time-of-flight~\cite{PhysRevA.80.033604, PhysRevA.98.043604}, useful for thermometry~\cite{Manz10, PhysRevA.104.043305}.
At equilibrium, the fluctuations are captured by the Bogoliubov theory~\cite{PhysRevA.67.053615} or the Luttinger liquid theory~\cite{Haldane81, Haldane94, Cazalilla04}, describing linearized 1D quantum fluids.
However, to capture the full quantum dynamics, non-linearized descriptions beyond the mean-field approach~\cite{RevModPhys.73.307} are needed.

The Generalized Hydrodynamics (GHD)~\cite{castro2016emergent, bertini2016transport} is one of such methods.
It is based on the assumption that a 1D system of bosonic atoms can be locally characterized by an equilibrated mixed state of the Lieb-Liniger model~\cite{LL1,LL2}.
Its solutions are parameterized in terms of quasi-particles, each characterized by its rapidity $\theta$, whose distribution at finite temperature is obtained via the thermodynamic Bethe ansatz (TBA)~\cite{YY1}.
GHD provides a coarse-grained description of dynamics by casting the coupled continuity equations for all the conserved quantities into a single transport equation for the quasi-particles.

Although GHD has been demonstrated to accurately describe the dynamics of 1D cold gas experiments~\cite{schemmer2019generalized, PhysRevLett.126.090602, malvania2020generalized, cataldini2021emergent}, the theory suffers from a few significant limitations:
Firstly, the quasi-particle distributions enable computation of local expectation values but do not provide information on the fluctuations. 
Secondly, the assumption of locality results in all spatially separated, equal-time connected correlations to vanish~\cite{doyon2018exact, 10.21468/SciPostPhysCore.3.2.016}.
Correlations of the phase in particular, which can be measured in interference experiments with split quasi-condensates~\cite{Hofferberth07}, have been instrumental in characterizing 1D systems~\cite{Gring1318, doi:10.1126/science.1257026, Schweigler17, RauerRecurrencies, Schweigler2021}.
In the present Letter we demonstrate a method to treat thermal fluctuations of a 1D bosonic degenerate gas within the GHD framework, thus advancing the capabilities of GHD to a new level, usually attainable by effective field theories.

The quantum collective variables of density and phase can be introduced using the bosonization procedure~\cite{Haldane81, Haldane94, Cazalilla04}.
Our idea is to start from a hydrostatic equilibrium for a 1D bosonic gas at zero temperature $T=0$~\footnote{See Supplemental Material for further details.}, which yields the unperturbed linear density profile $\varrho_0 (z)$.
The next step is to add density fluctuations $\delta \! \varrho (z)$ subject to the constraint $\int _{-\infty } ^\infty dz\, \delta \! \varrho (z)=0$. 
Knowing the perturbed local density $\varrho (z) = \varrho_0(z) + \delta \! \varrho (z)$ and the interaction parameter $c$, we can calculate the rapidity distribution for the ground state of the Lieb-Liniger model for each position $z$. 
Next, we recall the observation that a local, $z$-dependent Galilean boost of the Lieb-Liniger ground state results in the subsequent evolution describable in terms of conventional hydrodynamics~\cite{Doyon17a}.
Hence, we induce a phase shift $\phi (z)$ that gives rise  to the macrosciopic hydrodynamic velocity $V= (\hbar /m)\partial \phi /(\partial z)$.
At the microscopic description level, this means that the ground state distribution of rapidities at each $z$ experiences a shift (a Galilean boost) $\zeta (z) = \partial \phi (z)/(\partial z)$. 

The density and phase fluctuations can be quantized in a standard way~\cite{Haldane81, Cazalilla04}, yielding bosonic creation and annihilation operators for collective exciation quanta.
Note that these quanta do not correspond to individual particle-like or hole-like excitations of the Lieb-Liniger model~\cite{LL2}, but rather reflect the overcompleteness of the particle-hole basis that allows to express an excitation with a given momentum in infinitely many ways.
Herein lies the main difference between ours and other approaches~\cite{10.21468/SciPostPhys.8.3.048, PhysRevB.96.220302} including the quantum GHD ~\cite{PhysRevLett.124.140603, Ruggiero_2021}.
To distinguish excitations of the type considered in the present Letter from those used for the construction of the quantum GHD, we dub the former ones "boostons". 

\begin{figure}
    \centering
    \includegraphics[width = \columnwidth]{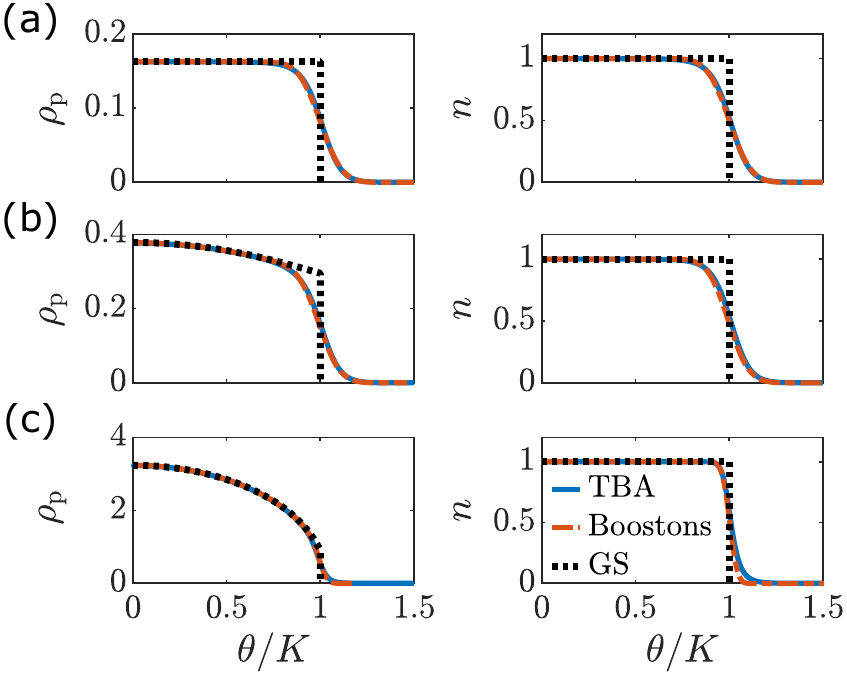}
    \phantomsubfloat{\label{fig:quasi-particle_distributions_a}}
    \phantomsubfloat{\label{fig:quasi-particle_distributions_b}}
    \phantomsubfloat{\label{fig:quasi-particle_distributions_c}}
    \vspace{-3\baselineskip}

    \caption{
    Quasi-particle distributions $\rho_\mathrm{p} (\theta)$ and occupations functions $n(\theta)$ at finite temperature obtained using the thermodynamic Bethe ansatz (TBA) and by sampling boostons from a thermal ensemble. For comparison, the ground state distribution obtained using the Bethe ansatz is also plotted. 
    (\textbf{a}) Strongly interacting regime (Tonks-Girardeau): $\gamma = 100$, $k_B T = 0.055 m v_s^2$.
    (\textbf{b}) Intermediately interacting regime: $\gamma = 1$, $k_B T = 0.15 m v_s^2$.
    (\textbf{c}) Weakly interacting regime (quasi-condensate): $\gamma = 0.01$, $k_B T = 0.25 m v_s^2$.}
    \label{fig:quasi-particle_distributions}
\end{figure}

Our first task is to show that the distribution of quasi-particles $\rho _\mathrm{p}(\theta )$ obtained from the TBA at thermal equilibrium can be reproduced, at least at low temperatures, by a thermal ensemble of boostons.
In the spatially uniform case at $T=0$, we have $\rho _\mathrm{p}(\theta ) = f(\gamma ,\theta /K)$ for $|\theta |\leq K$ and zero otherwise, where $\gamma =c/\varrho $ is the Lieb-Liniger parameter, $K$ is the "Fermi rapidity", and the function $f$ is introduced in Ref.~\cite{LL1}.
Adding excitations in the form of boostons means that at a given point $z$ the rapidity distribution changes to $\rho _\mathrm{p}(\theta )= f[\gamma ^\prime ,(\theta -\zeta )/K^\prime ]$ for $-K+\zeta \leq \theta \leq K+\zeta $ and 0 otherwise, where $\gamma ^\prime $ and $K	^\prime $ are defined for the new local density $\varrho ^\prime = \varrho +\delta \! \varrho $ and $\zeta =\partial \phi /\partial z$ is the local boost (statistically independent from the density fluctuation $\delta \! \varrho $).
Introducing the respective probability densities $w_{1,2}$, we obtain 
\begin{equation} 
    \rho _\mathrm{p}(\theta ) =\int _{-\infty }^\infty d \delta \!\varrho \, w_1(\delta \! \varrho )
    \int _{\theta -K^\prime }^{\theta +K^\prime }d\zeta \, w_2(\zeta )f[\gamma ^\prime ,(\theta -\zeta )/K^\prime ] .
    \label{dual.3} 
\end{equation} 
Since local density and phase fluctuations are superpositions of many thermally populated momentum modes~\cite{Note1}, which are statistically mutually independent for small thermal fluctuations, we can assume, according to the central limit theorem~\cite{Hudson63}, that the distributions $w_{1,2}$ are Gaussian with zero mean. 
We estimate the variances for the density and local boost fluctuations as 
$D_{\delta \! \varrho }=\langle : \delta \! \hat \varrho ^2:\rangle $ and $D_{\zeta }=\langle : \hat \zeta ^2:\rangle $, respectively, where we perform the second quantization of these fields and take thermal average of normally ordered operators in order to exclude zero-point oscillations.
Since boostons are bosons~\cite{Haldane81, Haldane94, Cazalilla04}, the result is 
\begin{equation} 
    D_{\delta \! \varrho }=\frac {\mathscr{K}}6\left( \frac {k_\mathrm{B}T}{\hbar v_\mathrm{s}K}\right) ^2K^2, 
    \quad  
    D_\zeta =\frac {\pi ^2}{6\mathscr{K}}\left( \frac {k_\mathrm{B}T}{\hbar v_\mathrm{s}K}\right) ^2K^2, 
\label{dual.4} 
\end{equation} 
where $v_\mathrm{s}$ is the sound velocity and $\mathscr{K}$ is the Luttinger liquid parameter~\cite{Haldane81, Cazalilla04, Haldane94, Note1} that has the asymptotics $\mathscr{K}\approx \pi /\sqrt \gamma $ for $\gamma \ll 1$ and $\mathscr{K}\rightarrow 1$ for $\gamma \rightarrow \infty $.
The assumption of normally distributed fluctuations with the variances given by Eq.~(\ref{dual.4}) holds only if $D_{\delta \! \varrho }$ is small compared to $\varrho ^2$, which corresponds to the low temperature range 
\begin{equation}
    k_\mathrm{B}T\lesssim mv_\mathrm{s}^2\,  {\min }(\gamma ^{-1/4},1).    
\label{dual.5} 
\end{equation} 
Estimations for higher temperatures are much more complicated and beyond the scope of the present Letter. 

\begin{figure}
    \centering
    \includegraphics[width = \columnwidth]{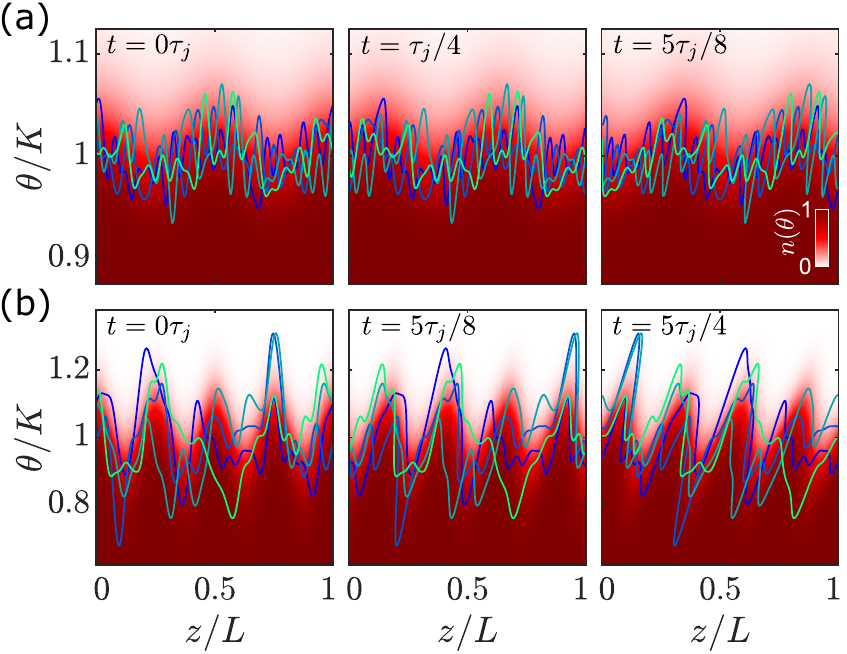}
    \phantomsubfloat{\label{fig:contour_evol_a}}
    \phantomsubfloat{\label{fig:contour_evol_b}}
    \vspace{-2\baselineskip}

    \caption{
    Evolution of occupation function $n$ parameterized as Fermi contours $\Gamma^+$ (colored lines) for a few thermal booston realizations with a single coherently excited density mode. For comparison, the evolution of the finite temperature occupation obtained using TBA is plotted as a 2d color plot underneath.
    The results are computed for a quasi-condensate (\textbf{a}) and a Tonks-Girardeau gas (\textbf{b}) and are plotted at select fractions of the mode period $\tau_j = 2 \pi /(k_j v_s)$.
    See main text for parameters used.
    }
    \label{fig:contour_evol}
\end{figure}

\begin{figure*}
    \centering
    \includegraphics[width = 0.9\textwidth]{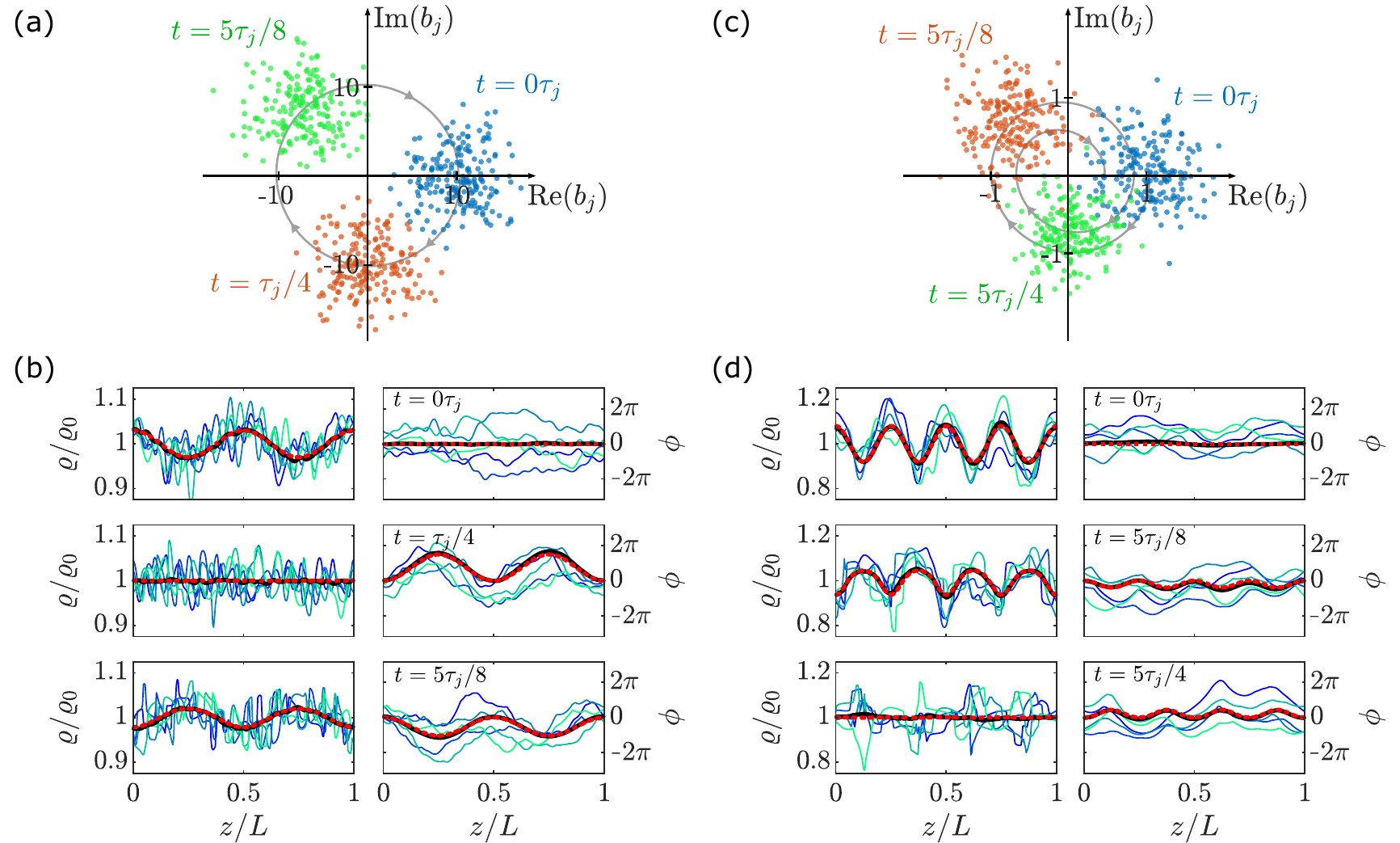}
    \phantomsubfloat{\label{fig:mode_evol_a}}
    \phantomsubfloat{\label{fig:mode_evol_b}}
    \phantomsubfloat{\label{fig:mode_evol_c}}
    \phantomsubfloat{\label{fig:mode_evol_d}}
    \vspace{-2\baselineskip}
    
    \caption{
    Evolution of states with thermally sampled boostons and a single density mode coherently excited. 
    (\textbf{a}) Bogoliubov mode operators $b_j$ of Eq.~(\ref{eq:mode_operators}) at select fractions of the mode period $\tau_j = 2 \pi /(k_j v_s)$. Here, the $j= \pm 2$ mode of a quasi-condensate is coherently excited and plotted. The grey line indicates the evolution of $\langle b_j \rangle$ up to time $t = \tau_j$.
    (\textbf{b}) Corresponding density $\varrho (z)$ and phase profiles $\phi (z)$. The thin lines are examples of profiles for a few individual realizations, while the thick black line is the expectation value computed via ensemble averaging. For comparison, expectation values obtained using GHD with TBA initial state are plotted as dashed lines.
    (\textbf{c, d}) Similar plots here for the coherently excited $j= \pm 4$ density mode in a Tonks-Girardeau gas. The evolution of $\langle  b_j \rangle$ is plotted up to time $t = 2 \tau_j$.
    }
    \label{fig:mode_evol}
\end{figure*}

We can numerically test the approximation of Eqs.~(\ref{dual.3},~\ref{dual.4}) by 
expanding the density and phase fluctuation fields in eigenmodes, sampling boostons 
from Bose-Einstein distribution and then finding the local quasi-particle distributions for the
resulting values of $K'$ by solving the ground-state Bethe ansatz~\cite{Note1}.
Averaging over many such realizations yields the distributions plotted in Fig.~\ref{fig:quasi-particle_distributions}, which agree very well with the TBA results.
Equivalently, a state can be parameterized through the occupation function $n(\theta) = \rho_\mathrm{p} (\theta) / \varpi (\theta )$, where $\varpi (\theta )$ is the density of states~\cite{YY1}.
According to the TBA the Lieb-Liniger rapidities exhibit fermionic statistics, whereby the $T=0$ occupation reads $ n(\theta) = 1$ for $|\theta |\leq K$ and 0 otherwise, i.e., a Fermi sea in the rapidity space.
At finite temperature the edge of the Fermi sea starts to "melt", as higher rapidities become thermally occupied, thus creating a smooth transition between the completely filled and completely empty rapidity states. 
From the normalised distribution of sampled Fermi momenta $w (K')$, we can approximate the occupation function near the right Fermi momentum following $n(\theta) = 1 - \int_{0}^{\theta} \mathrm{d}K' \, w(K') $.
Again, we find good agreement with the results of TBA, as seen in Fig.~\ref{fig:quasi-particle_distributions}.
Thus, at low temperatures, thermal distributions of the thermodynamic Bethe ansatz can be reproduced by a thermal ensemble of boostons.

Next, we seek to demonstrate how local expectation values obtained form a TBA thermal state propagated according to GHD can be reproduced by thermally sampled boostons propagated following the same GHD principles~\cite{Note1}.
Owing to local fluctuations, a single realization in the booston picture is given by a locally perturbed zero-temperature state.
Thus, the occupation function of a single realization reads $ n(\theta, z) = 1$ for $\Gamma^- (z) \leq \theta \leq\Gamma^+ (z)$ and 0 otherwise, where $\Gamma^\pm (z)= \pm K' (z) + \zeta (z)$ is the boosted Fermi contour.
Time-dependent expectation values are then computed via ensemble averaging over a number of realization individually propagated according to GHD.
Crucially, this approach accounts for the contribution of the fluctuations to the dynamics, unlike most low-energy effective field theories~\cite{PhysRevA.67.053615} or the quantum GHD~\cite{PhysRevLett.124.140603} which linearize the equations of motion for the fluctuations around a background state.

For the demonstration, we study the setup of a thermal state with a single, coherently excited density mode, inspired by recent experimental protocols~\cite{cataldini2021emergent}.
We employ periodic boundary conditions and sample boostons for the 40 lowest momentum modes, whose thermal populations follow the Bose-Einstein distribution. 
We perform the numerical experiment for two different systems: (\textit{i}) a quasi-condensate with Lieb-Liniger parameter $\gamma = 0.01$, temperature $k_B T = 0.25 m v_s^2$, and the $j = \pm 2$ density mode with momentum $k_j = 2 \pi j /L$ initially coherently excited, and (\textit{ii}) a Tonks-Girardeau gas with $\gamma = 100$, $k_B T = 0.055 m v_s^2$ and the $j = \pm 4$ mode excited.

First, we simulate the GHD evolution~\cite{10.21468/SciPostPhys.8.3.041} of thermal states obtained using the thermodynamic Bethe ansatz, whose occupation function we plot in Fig.~\ref{fig:contour_evol_a} and \ref{fig:contour_evol_b} for the quasi-condensate and Tonks-Girardeau gas, respectively.
Note that only the rapidities near the corresponding zero-temperature Fermi momentum $K$ are shown.
The spatial variation of the occupation function reflects the coherently excited density mode.
For comparison, the Fermi contours $\Gamma^+(z)$ of a few propagated booston realizations are plotted on top of the TBA occupation functions. 
The contours follow in general the shape of the coherent mode, 
however, it is slightly obscured by the thermal fluctuations.

\begin{figure}
    \centering
    \includegraphics[width = 0.9\columnwidth]{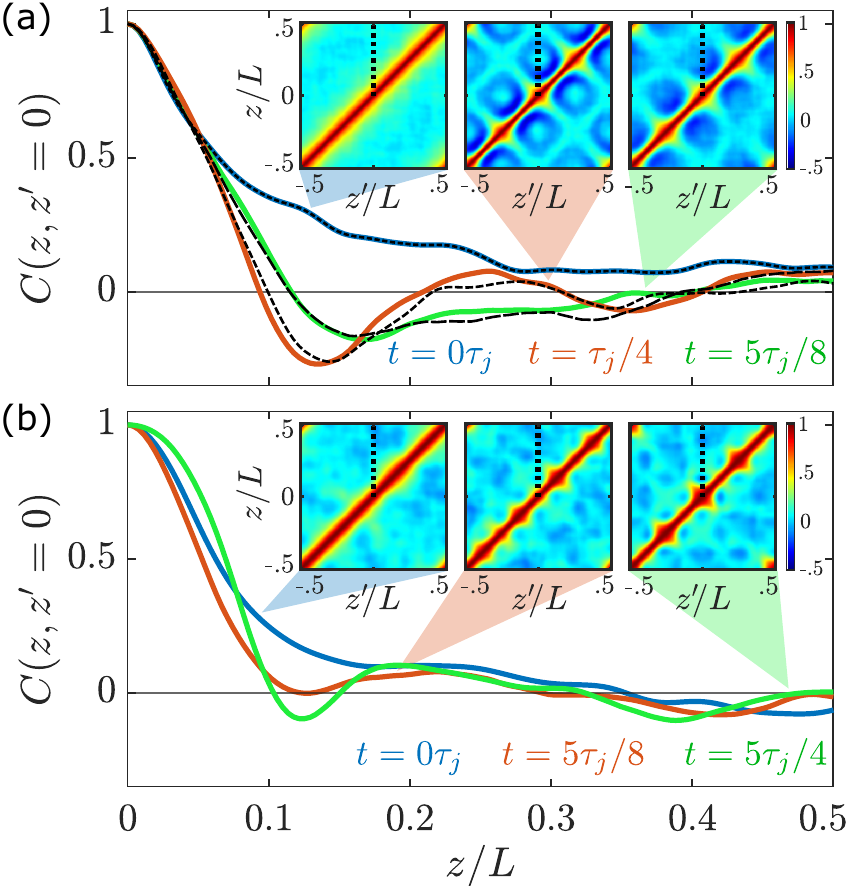}
    \phantomsubfloat{\label{fig:correlation_evol_a}}
    \phantomsubfloat{\label{fig:correlation_evol_b}}
    \vspace{-2\baselineskip}

    \caption{
    Two-point phase correlation function $C(z,0)$ of Eq.~(\ref{eq:twopoint_phase_corr}) plotted at the times indicated on the plot for (\textbf{a}) the quasi-condensate and (\textbf{b}) the Tonks-Girardeau gas. See main text for parameters used.
    The dashed lines in (\textbf{a}) are obtained from GPE calculations with the same initial state (short-dashed line for $t=\tau _j /4$, long-dashed line for $t=5\tau _j /8$).
    The insets show the full correlation functions $C(z,z')$, where the corresponding $C(z,0)$ curves are indicated by a dotted line.
    }
\end{figure}

To better visualize the evolution of the coherently excited mode, we extract the density fluctuations $\delta \varrho(z)$ and phase $\phi(z)$ for each booston realization following evolution.
From these we can compute the Bogoliubov mode operators~\cite{PhysRevA.67.053615}
\begin{equation}
    b_j = \int_{-\infty}^{\infty} \mathrm{d}z \: \frac{1}{\sqrt{4 \varrho_0}} \left[ \bar f_j^- \delta \varrho (z) + i \sqrt{\varrho_0} \bar f_j^+ \phi(z) \right] \; ,
    \label{eq:mode_operators}
\end{equation}
where $\bar f_j^\pm$ are the complex conjugated mode functions of the $j$'th mode for periodic boundary conditions.
In Figs.~\ref{fig:mode_evol_a} and \ref{fig:mode_evol_c} we plot the mode operators of the excited mode for the two systems.
By virtue of our sampling, the distribution of $b_j$ corresponds to a coherent state, whose variance is determined by the temperature, and which is initially offset on the real axis indicating the coherent population in the density quadrature. 
Following dynamics the coherent amplitude $\langle  b_j \rangle$ decreasing over time, as the excited mode decays into higher modes.
For the given scenario, the decay follows from the fully interacting nature of the system achieved by propagating the full state according to GHD.

Next, we compute the expectation values of the phase and density through ensemble averaging, which are plotted as black lines in Fig.~\ref{fig:mode_evol_b} and \ref{fig:mode_evol_d}. 
For reference, a few profiles of individual realizations are plotted as thin, colored lines.
Initially, the coherently exited mode ($j=2$ mode for the quasi-condensate and $j=4$ mode for the Tonks-Girardeau gas) can be seen clearly in the density profile, while the fluctuating phase profiles average to zero.
During the evolution the excited mode rotates into the phase quadrature, where it is at its maximum after a propagation time of $t = \tau_j/4$, with $\tau_j = 2 \pi /(k_j v_s)$ being the period of the $j$'th mode.
Comparing the expectation values to TBA results (plotted as dashed lines) we find the two approaches to agree well, both in the quasi-condensate and the Tonks-Girardeau regime.
Thus, the results plotted in Figs.~\ref{fig:contour_evol} and \ref{fig:mode_evol} demonstrate that the approximation of thermal states through booston excitations remains valid following the GHD evolution.

Finally, we wish to demonstrate how the booston excitations enable us to compute correlation functions. 
Thus, from the phase profiles of the individual booston realizations we calculate the two-point phase correlation function
\begin{equation}
    C(z,z') = \langle e^{i (\phi(z) - \phi(z'))} \rangle \; ,
\label{eq:twopoint_phase_corr}
\end{equation}
which is routinely measured in experiments with ultracold atomic gases.
In Fig.~\ref{fig:correlation_evol_a} we plot the resulting correlation functions for the quasi-condensate, while Fig.~\ref{fig:correlation_evol_b} shows the results in the Tonks-Girardeau regime. 
Initially, at time $t = 0$, the phase correlations are completely thermal, as the coherent excitation is confined to the density quadrature.
However, as the coherent excitation rotates into the phase quadrature, a pattern in the shape of the excited mode emerges in the full two-point correlation function $C(z,z')$.

For comparison we show $C(z,0)$ in the quasi-condensate regime, when evolving the initial states with the 1d Gross-Pitaevskii equation (GPE) for the same value of $\gamma $ and the stochastic initial conditions sampling the fluctuations for the same temperature.
We find good agreement between the GHD and GPE evolution.
Since the mean-field method works well for quasi-condensates, this agreement indicates the correctness of both  our theoretical approach and its numerical implementation.

We presented a new method to treat fluctuations within the GHD framework.
Representing the quasi-particle distribution through an ensemble of states sampled within the bosonic 'booston' basis, our method enables the calculation of equal-time correlations evolving the ensemble with the standard GHD equations.
In particular, our results extend beyond the linearized regime of quantum GHD~\cite{PhysRevLett.124.140603, Ruggiero_2021} by taking into account backaction of the fluctuations.
The presented GHD method enables to calculate correlations for arbitrary Lieb-Liniger parameters $\gamma$, with good agreement to predictions of the GPE within the weakly interacting quasi-condensate regime.
Higher-order equal-time correlations provide a new precise method to compare predictions of GHD to other numerical methods or experimental results.

%% file: supplemental_material.tex
\section{Derivation of hydrodynamic equations from GHD} 

The GHD evolution equation reads as 
\begin{equation} 
    \frac \partial {\partial t} \rho _\mathrm{p} +\frac \partial {\partial z}\left( v^\mathrm{eff} \rho _\mathrm{p}\right)  =0
\label{GHD.1} 
\end{equation} 
and is supplemented by two equations 
\begin{equation} 
    v^\mathrm{eff} (\theta ) = \frac {\hbar \theta }m +\int _{-\infty }^\infty d\theta ^\prime \, \frac {2 c \rho _\mathrm{p} (\theta ^\prime )}{c^2+(\theta -\theta ^\prime )^2} \left[ v^\mathrm{eff} (\theta ^\prime )-v^\mathrm{eff} (\theta )\right]  \; , 
\label{GHD.2} 
\end{equation} 
\begin{equation} 
    \varpi (\theta ) =\frac 1{2\pi }+\frac 1\pi \int _{-\infty }^\infty d\theta ^\prime \, \frac c{c^2+(\theta -\theta ^\prime )^2} \rho _\mathrm{p} (\theta ^\prime ) \; , 
\label{GHD.3} 
\end{equation}   
where $m$ is the mass of the atoms, $c$ is the coupling strength, and the density of states 
\begin{equation} 
    \varpi (\theta ) = \rho _\mathrm{p} (\theta ) + \rho _\mathrm{h} (\theta )
\label{GHD.4} 
\end{equation} 
is the sum of the densities of quasi-particles (p) and holes (h). 
We introduce the mean hydrodynamic velocity 
\begin{equation} 
    V =\frac 1\varrho \int _{-\infty }^\infty d\theta \, \frac {\hbar \theta }m \rho _\mathrm{p} (\theta ), 
\label{GHD.5}
\end{equation} 
where
\begin{equation}
    \varrho =\int _{-\infty }^\infty d\theta \, \rho _\mathrm{p} (\theta )
\end{equation}
is the linear density of atoms. It follows from Eq. (\ref{GHD.2}) that 
\begin{equation} 
    V =\frac 1\varrho \int _{-\infty }^\infty d\theta \, v^\mathrm{eff} (\theta )\rho _\mathrm{p} (\theta ). 
\label{GHD.6}
\end{equation} 
Consider the rapidity distribution with a co-ordinate-dependent Galilean boost
\begin{subequations}
\begin{align}
    \rho _\mathrm{p} (\theta ) &= \bar \rho _\mathrm{p} (\theta -\zeta ) \; ,\\
    v^\mathrm{eff} (\theta ) &= \bar v^\mathrm{eff} (\theta -\zeta )+V \; , 
    \label{GHD.7} 
\end{align}
\end{subequations}
where 
\begin{equation}
    \zeta (z) = m V(z)/\hbar 
\end{equation}
is the rapidity characterizing the local boost and the barred quantities describe the system in the co-moving frame of reference. 
By definition, 
\begin{equation}
    \int _{-\infty }^\infty d\theta \, \bar v^\mathrm{eff} (\theta )\bar \rho _\mathrm{p} (\theta )=0 \; .    
\end{equation}
The local density $\varrho (z)$ also can depend on the co-ordinate.   
Integrating Eq. (\ref{GHD.1}) over rapidities, we obtain the continuity equation 
\begin{equation} 
    \frac \partial {\partial t}\varrho +\frac \partial {\partial z}(\varrho V)=0 \; . 
\label{GHD.8} 
\end{equation} 
Multiplying Eq. (\ref{GHD.1}) by $\hbar \theta /m$ and then integrating over $\theta $, we obtain the momentum balance equation 
\begin{equation}
    \frac \partial {\partial t}(\varrho V)+ \frac \partial {\partial z}\int _{-\infty }^\infty d\theta \, \frac {\hbar \theta }m v^\mathrm{eff}(\theta )\rho _\mathrm{p}(\theta ) + \frac \varrho m \frac {\partial U}{\partial z}=0 \; ,    
\end{equation}
which can be further transformed, using Eq.~(\ref{GHD.7}), to 
\begin{equation} 
    \frac {\partial V}{\partial t}+V\frac {\partial V}{\partial z}=-\frac 1{m\varrho } \frac {\partial \mathcal{P}}{\partial z}-\frac 1m \frac {\partial U}{\partial z} \; ,  
\label{GHD.9} 
\end{equation} 
where 
\begin{equation} 
    \mathcal{P}=\int _{-\infty }^\infty d\theta \, \hbar \theta v^\mathrm{eff}(\theta )\rho _\mathrm{p} (\theta )
\label{press} 
\end{equation} 
is the 1D pressure. Eqs. (\ref{GHD.8}, \ref{GHD.9}) have a form of the standard equations of macroscopic hydrodynamics, supplemented by the equation of state (\ref{press}).
These equations are essentially classical: the quantum potential 
$-[\hbar ^2/(2m \sqrt \varrho )] \partial ^2 \sqrt \varrho /(\partial z^2)$ does not appear (added to $U$) in the r.h.s. of Eq. (\ref{GHD.9}), because the GHD is an essentially local theory. 
Using Eq. (\ref{GHD.3}), we can rewrite Eq. (\ref{GHD.2}) for the effective velocity in the co-moving frame as 
\begin{equation} 
    \bar \varpi (\theta ) \bar v^\mathrm{eff}(\theta ) = \frac {\hbar \theta }{2\pi m} +\frac 1\pi \int _{-\infty }^\infty d\theta \, \frac c{c^2 +(\theta -\theta ^\prime )^2} \bar v^\mathrm{eff}(\theta )\bar \rho _\mathrm{p}(\theta ) \; . 
\label{GHD.10}  
\end{equation}

\section{Sampling of boostons}

Fluctuations of the local Fermi points in the form of boostons translate into local fluctuations of density $\delta \varrho$ and boost $\zeta$, and vice versa. The density and boost (phase) fluctuations are quantized \cite{Haldane81,Cazalilla04}
\begin{align}
    \delta\varrho (z) &= \sqrt{ \varrho_0 } \sum_j \left[ f_j^+ (z)  b_j + \mathrm{H.c.} \right] \label{eq:density_fluct_mode} \\
    \zeta (z) &= \frac{1}{\sqrt{ 4 \varrho_0 }} \sum_j \left[ k_j f_j^- (z)  b_j + \mathrm{H.c.} \right] \; , \label{eq:boost_fluct_mode}
\end{align}
with the creation and annihilation operators of the collective (mode) excitations obeying the usual bosonic commutation relations $[ b_i ,  b_j^\dagger] = \delta_{ij}$.
For simplicity the background density $\varrho_0$ is assumed to homogeneous, while coherent variations in density and boost added as part of the mode operators. 
The mode functions $f_j^\pm$ appearing in Eqs. (\ref{eq:density_fluct_mode}, \ref{eq:boost_fluct_mode}) are normalized as
\begin{equation}
    \frac{1}{2} \int \mathrm{d}z \: \left[ \bar f_j^+ f_j^- + f_j^+ \bar f_j^- \right] = 1 \; ,
\end{equation}
where the bar here denotes complex conjugate. For the majority of this work we concern ourselves with periodic boundary conditions, in which case the mode functions take the form
\begin{equation}
    f_j^\pm (z) = \frac{1}{\sqrt{L}} \left( \frac{\epsilon_j}{E_j} \right)^{\mp 1/2} \mathrm{e}^{ i k_j z } \; ,
\end{equation}
with $k_j = 2 \pi j / L$, $E_j = \hbar^2 k_j^2 / 2 m$, and $\epsilon_j$ being given by the Bogoliubov spectrum
\begin{equation}
    \epsilon_j = \sqrt{E_j (E_j + 2 m v_s^2)} \; .     
\end{equation}
We assume the population of the individual modes to be follow the Bose-Einstein distribution $n_j = 1/(\mathrm{e}^{\epsilon_j / k_B T} - 1)$. 
To achieve a thermal ensemble of boostons, the mode operators are sampled following
\begin{equation}
     b_j = \alpha_j \mathrm{e}^{ i \varphi_j } + \sqrt{n_j} \frac{ X_1 + i  X_2}{\sqrt{2}} \; ,
    \label{eq:mode_operator_sample}
\end{equation}
where $X_1$ and $X_2$ are sampled from indepedent, Gaussian distributions with zero mean and unit variance.
The first term of Eq. (\ref{eq:mode_operator_sample}) accounts of any coherent population of the mode, shifting the expectation value of the creation operator $\langle  b_j \rangle = \alpha_j \langle \mathrm{e}^{i  \varphi_j} \rangle$.
Given the thermally sampled creation operators, the corresponding density and boost fluctuations are computed via Eqs. (\ref{eq:density_fluct_mode}, \ref{eq:boost_fluct_mode}), which in turn are translated into fluctuations in the Fermi edge (boostons) by locally solving the (zero-temperature) Lieb-Liniger equations.
To this end, significant computational advantage can be achieved by solving the equations beforehand for a range of interaction strengths $\gamma$ and then tabulating the results.

\section{Numerical implementation of zero-temperature GHD with boostons}
The zero-temperature GHD (+boostons) simulations in this work employ the algorithm originally featured in Ref. \cite{Doyon17a}, while TBA and finite temperature GHD calculations were performed using the iFluid framework \cite{10.21468/SciPostPhys.8.3.041}.

\begin{figure}  
    \centering
    \includegraphics[width=78mm]{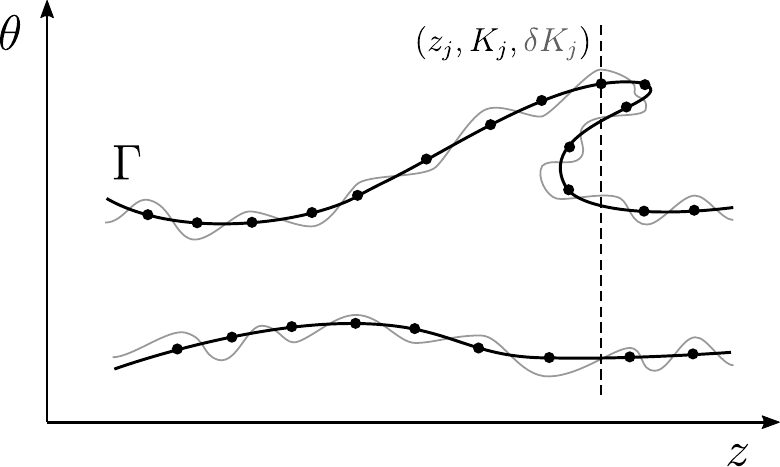} 
    \caption{
    Illustration of algorithm for numerically propagating the state parameterized as the Fermi contour $\Gamma$, which in turn is discretized as a set of points (see text for details). The coherent part of the contour is drawn in black, while the grey lines mark the contours after adding incoherent fluctuations $\delta K$ in the form of boostons.}
    \label{fig:algorithm}
\end{figure}

For the GHD simulations, rather than propagating the quasi-particle distribution $\rho_\mathrm{p} (\theta)$, it is numerically more convenient to work with the occupation function $n (\theta)$, which for a zero-temperature state with Fermi rapidity $K$ is given by
\begin{equation}
    n(\theta) =
    \begin{cases}
        1, \qquad \text{for } -K \leq \theta \leq K \\
        0, \qquad \text{otherwise, }
    \end{cases}
\end{equation}
thus realizing a Fermi sea.
In the presence of multiple local Fermi seas $K_1^- < K_1^+ < K_2^- < K_2^+ < \ldots$, the occupation function will assume the value 1 between the rapidity pairs $(K_i^- , K_i^+)$ and 0 anywhere else. 
It is thus sufficient to encode the state only by the local Fermi points, whose displacement following dynamics is given by
\begin{equation}
    \partial_t K_i^\pm + v_{\{ K \}}^\mathrm{eff} (K_i^\pm) \; \partial_z K_i^\pm = 0 \; .
    \label{eq:Fermi_point_propagation}
\end{equation}
Here the effective velocity is computed using
\begin{equation}
    v_{\{ K \}}^\mathrm{eff} (\alpha) = \frac{ \mathrm{id}_{\{ K \}}^\mathrm{dr} (\alpha)}{ 1_{\{ K \}}^\mathrm{dr} (\alpha)} \; ,   
    \label{eq:veff_contour}
\end{equation}
which is an equivalent expression than that of Eq. (\ref{GHD.2}). In Eq. (\ref{eq:veff_contour}), $\mathrm{id}(\alpha)$ is the identity function and the dressing operation is defined as
\begin{equation}
    f_{\{ K \}}^\mathrm{dr} (\alpha) = f(\alpha) + \sum_{i=1}^\mathcal{N} \int_{K_i^-}^{K_i^+} \mathrm{d}\theta \frac{2 c}{ c^2 + (\alpha - \theta)^2}  f_{\{ K \}}^\mathrm{dr} (\theta) \; ,
\end{equation}
where $\mathcal{N}$ is the number of local Fermi seas.
Following Ref. \cite{Doyon17a}, we can encode the time- and space-dependent state $n(\theta,z,t)$ as a Fermi contour $\Gamma$ containing all the Fermi points. Discretizing the contour yields a set of points ($z_j (t), K_j (t)$), whose displacement after a small time step $\delta t$ is
\begin{equation}
    z_j (t+\delta t) = z_j (t) + \delta t \: v_{\{ K \}}^\mathrm{eff} (K_j) \; .
    \label{eq:dispacement}
\end{equation}
Note, in the absence of any acceleration of the quasi-particles, which is the case in Eq. (\ref{eq:Fermi_point_propagation}), all rapidities are conserved, whereby all $K_j$ remain constant.
The numerical propagation of the state follows from a simple algorithm: For each point $j$ in the contour $\Gamma$, find all local Fermi seas by searching for intersections between the vertical line at $z = z_j$ and the contour itself (see Fig. \ref{fig:algorithm}). Next, compute the effective velocity at each point using Eq. (\ref{eq:veff_contour}). Finally, displace each point of the contour according to Eq. (\ref{eq:dispacement}) and repeat.

\section{Relaxation of coherently excited modes}

\begin{figure}  
    \centering
    \includegraphics[width=78mm]{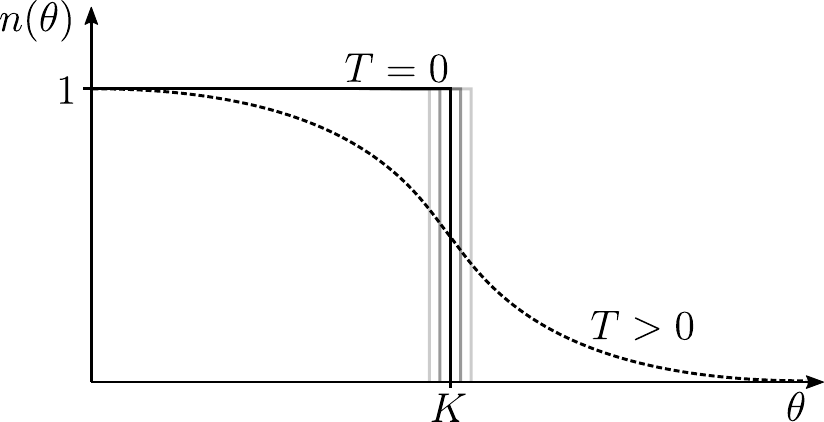} 
    \caption{ 
    Illustration of occupation function $n (\theta)$, which following the thermodynamic Bethe ansatz is given by a Fermi distribution, for zero temperature (solid line) and finite temperature (dashed line).
    At finite temperature the Fermi point (rapidity) $K$, and thereby the sound velocity following Eq. (\ref{eq:sound_velocity}), is ill-defined.
    For small enough temperatures, thermal fluctuations can be treated as local fluctuations of the Fermi point in the form of boostons, thus leading to local fluctuations of the sound velocity.}
    \label{fig:fermi_edge_illustration}
\end{figure}

In Bogoliubov's theory for the interacting Bose gas, the equation of motion are linearized around a stationary solution, thus neglecting any back-action of the fluctuations.
Hence, the Hamiltonian of the fluctuations is given by a sum of uncoupled harmonic oscillators, whereby the populations of all modes are conserved during evolution. 
The linearization step is reasonable for very low temperatures and small coherent populations of the modes, however, beyond this low-energy regime the back-action of fluctuations must be taken into account, which inevitably will lead to relaxation of excited modes.
Within the thermodynamic Bethe ansatz the full thermodynamic state is characterized by the quasi-particle distribution, while thermal effects on dynamics are implicitly accounted for in GHD through the effective velocity.
An example of this can be found in Ref.~\cite{cataldini2021emergent}, where GHD was employed to describe the relaxation of a single excited density mode.
Although the excited mode was well-defined in $k$-space, its corresponding rapidity distribution spanned over a range of rapidities by virtue of finite temperature.
During evolution of the mode, its different rapidity components propagated at slightly different velocities, resulting in a gradual dephasing of the mode.
While GHD does provide an incredibly powerful framework for treating finite temperature dynamics, the exact role of temperature is rather opaque within the TBA.
As we have demonstrated in this work, the TBA quasi-particle distributions (at lower temperatures) can be described by a thermal ensemble of boostons.
Hence, employing this framework we can study the apparent dephasing of a single excited density mode in greater detail.

First, consider the microscopic definition of the sound velocity $v_s \equiv \mathrm{lim}_{k \to 0} \frac{\partial \varepsilon}{\partial k} $, which is derived from the excitation spectrum with energy $\varepsilon (k)$ and momentum $k$.
For a fermionic system near the ground state, all excitations are limited to momenta near the Fermi momentum. In the Bethe ansatz of the Lieb-Liniger model, whose quasi-particle excitations are fermiones, the sound velocity therefore reads~\cite{korepin_bogoliubov_izergin_1993}
\begin{equation}
    v_s  ={  \frac{\partial_\theta \varepsilon (\theta)}{\partial_\theta k(\theta)}  \bigg\vert } _{\theta= K} \equiv v^{\mathrm{eff}} (K) \; .
    \label{eq:sound_velocity}
\end{equation}
Thus, for a zero-temperature state of the Lieb-Liniger model (with only a single local Fermi sea) the sound velocity is equal to the effective velocity at the Fermi point $K$. 
Accounting for thermal density fluctuations leads to local fluctuations in the the Fermi point $\delta K$, and thereby local fluctuations in the sound velocity.
In the presence of phase fluctuations, equivalent to a local boost of the ground state distributions, we obtain a fluctuating advection velocity $v = v_s + \hbar \zeta / m$ in the laboratory frame.
These fluctuations are implicitly accounted for in the thermodynamic Bethe ansatz (TBA) by "melting" the edge of the Fermi sea of rapidities, as illustrated in Fig.~\ref{fig:fermi_edge_illustration}. 
At low enough temperatures, averaging over thermally sampled booston fluctuations reproduces the TBA equilibrium distributions, as demonstrated in the main text. 
Hence, at low temperatures the two approaches should predict the same relaxation rate of a single excited mode.

\begin{figure*}  
    \centering
    \includegraphics[width=\textwidth]{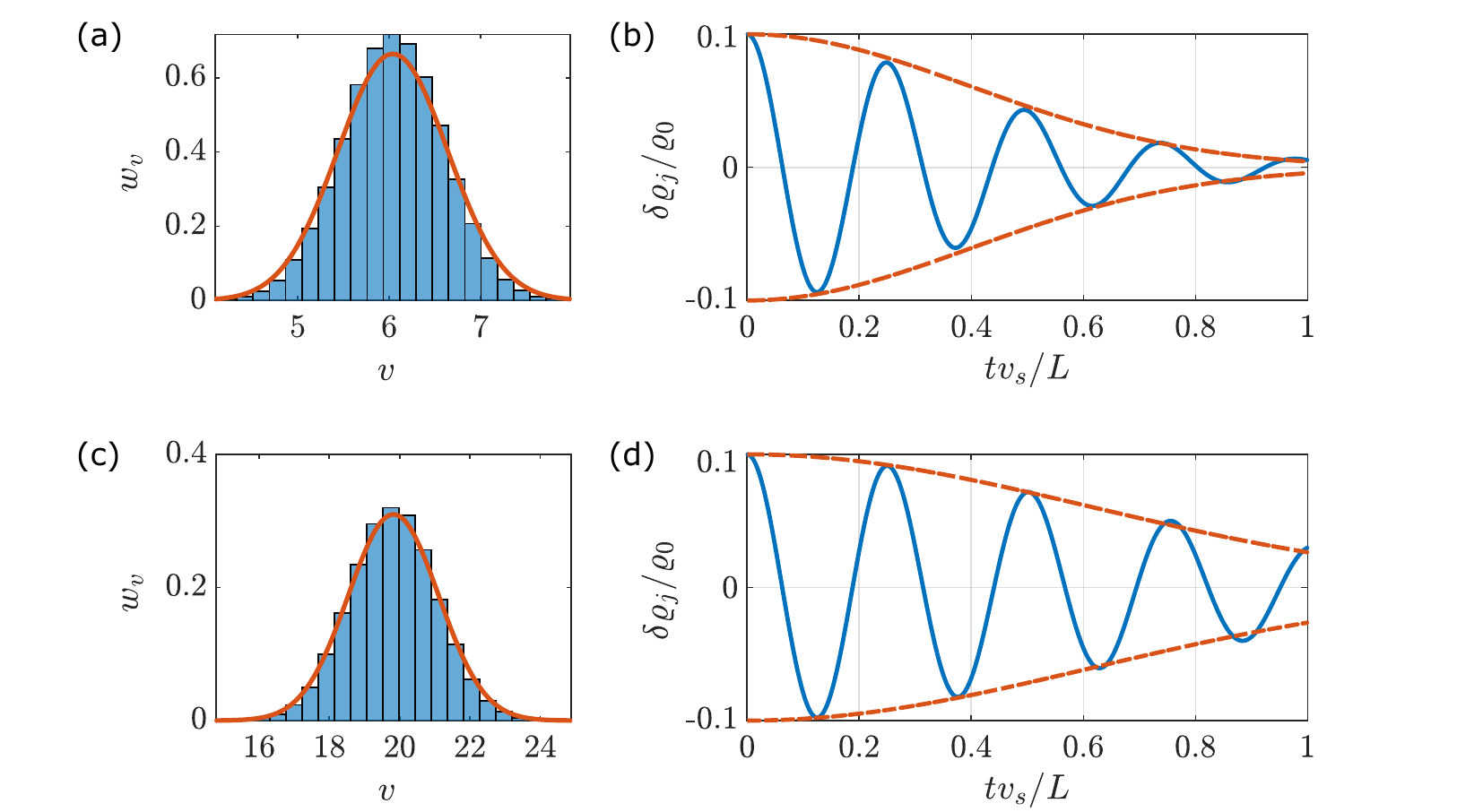} 
    \phantomsubfloat{\label{fig:mode_relaxation_a}}
    \phantomsubfloat{\label{fig:mode_relaxation_b}}
    \phantomsubfloat{\label{fig:mode_relaxation_c}}
    \phantomsubfloat{\label{fig:mode_relaxation_d}}

    \vspace{-2\baselineskip}
    
    \caption{
    (\textbf{a}) Sampled distribution of advection velocities for a Tonks-Girardeau gas with interaction $\gamma = 100$ and temperature $k_B T = 0.11 m v_s^2$. For comparison, a Gaussian distribution with variance $D_{\delta v}^{\mathrm{TG}}$ of Eq. (\ref{eq:variance_sound_TG}) is plotted.
    (\textbf{b}) Corresponding relaxation of a single, coherently excited density mode. The dynamics of the $j = 4$ mode is computed using GHD with TBA initial state (blue, solid curve). Its relaxation is compared to the damping term $\mathrm{e}^{ - \frac{1}{2} D_{\delta v} k_j^2 t^2}$ (red, dashed lines) of Eq. (\ref{eq:mode_damping}).
    \textbf{(c)} Sampled distribution of advection velocities for a quasi-condensate with interaction $\gamma = 0.01$ and temperature $k_B T = 0.25 m v_s^2$. For comparison, a Gaussian distribution with variance $D_{\delta v}^{\mathrm{QC}}$ of Eq. (\ref{eq:variance_sound_QC}) is plotted.
    \textbf{(d)} Corresponding relaxation of the $j=4$ density mode.}
    \label{fig:mode_relaxation}
\end{figure*}

The local advection velocity is dependent on the local density and boost. Following the central limit theorem the local fluctuations in density and boost are Gaussian. Further, we assume density and boost fluctuations to be independent, thus making fluctuations in advection velocity Gaussian as well
\begin{equation}
    w_{v}(\delta v) = \frac{\exp \left[ - \delta v^2 / (2 D_{\delta v})  \right]}{\sqrt{2 \pi D_{\delta v}}} \; .
\end{equation}
Consider the evolution of a single coherent density mode following Bogoliubov theory
\begin{equation}
    \delta\varrho_j (z,t) = \sqrt{ \varrho_0 } \left[ f_j^+ (z) \mathrm{e}^{-i \epsilon_j t / \hbar}   \alpha_j + \mathrm{H.c.} \right] \; ,
\end{equation}
where $\alpha_j = \langle b_j \rangle$ is the coherent amplitude. 
For simplicity we restrict ourselves to the phononic branch of the Bogoliubov spectrum where $\epsilon_j \approx \hbar k_j v_s$.
To account for local fluctuations in the advection velocity we let $v = v_{s0} + \delta v$, where $v_{s0}$ is the sound velocity computed with respect to the homogeneous background density $\varrho_0$.
Averaging over the fluctuations, we obtain the following expression for the evolution of the mode
\begin{align}
    \delta\varrho_j (z,t) &= \int_{-\infty}^{\infty} \mathrm{d}\delta v \:  w_{v}(\delta v) \sqrt{ \varrho_0 }
    \nonumber \\ 
    &\qquad \qquad \times \left[ f_j^+ (z) \mathrm{e}^{-i k_j (v_{s0} + \delta v)  t } \alpha_j + \mathrm{H.c.} \right] \nonumber \\
    &= \sqrt{ \varrho_0 } \left[ f_j^+ (z) \mathrm{e}^{-i v_{s0} t   }  \alpha_j + \mathrm{H.c.} \right] \mathrm{e}^{ - \frac{1}{2} D_{\delta v} k_j^2 t^2  } \; .
    \label{eq:mode_damping}
\end{align}
Indeed, in the presence of fluctuations the mode evolves as a damped oscillation, whose damping exponent scales quadratic with momentum and time. 
For comparison, in Ref.~\cite{cataldini2021emergent} the observed dynamics was fitted with damping exponents scaling with $\propto k_j^{3/2}, t^{3/2}$.
However, from GHD simulations the exact power was found to be slightly temperature dependent, with lower temperature realizations tending towards quadratic scaling.
Thus, the behavior of Eq.~(\ref{eq:mode_damping}) should be reproduced by GHD (with TBA initial states) in the low temperature limit. 

To obtain the scaling of the damping with temperature, we must compute the variance of the advection velocity $D_{\delta v}$.
This can be achieved analytically in the Tonk-Girardeau and quasi-condensate regime, where exact expressions for the sound velocity are known.
Hence, in the Tonk-Girardeau (TG) regime we have
\begin{align}
    v^{\mathrm{TG}} &= \frac{\pi \hbar \varrho}{m} + \frac{\hbar \zeta}{m} \\
    D_{\delta v}^{\mathrm{TG}} &= \left(\frac{\pi \hbar}{m}\right)^2 D_{\delta \varrho} + \left(\frac{ \hbar}{m}\right)^2 D_{\zeta} \label{eq:variance_sound_TG} \; ,
\end{align}
where the variances of the density and boost fluctuations are given in Eq. (\ref{dual.4}).
Likewise, in the quasi-condensate (QC) regime, the variance reads
\begin{align}
    v^{\mathrm{QC}} &= \frac{\hbar}{m} \sqrt{ c \varrho}  + \frac{\hbar \zeta}{m} \\
    D_{\delta v}^{\mathrm{QC}} &= \left(\frac{\hbar c }{m \sqrt{c \varrho}}\right)^2 D_{\delta \varrho} + \left(\frac{ \hbar}{m}\right)^2  D_{\zeta} \label{eq:variance_sound_QC} \; .
\end{align}
Figs.~\ref{fig:mode_relaxation_a} and \ref{fig:mode_relaxation_c} depict histograms of advection velocities calculated via Eq. (\ref{eq:sound_velocity}) for a state with thermally sampled boostons.
The sampled velocities are compared to the analytic distributions of Eqs. (\ref{eq:variance_sound_TG}, \ref{eq:variance_sound_QC}) and a good agreement is found.
Further, in Figs.~\ref{fig:mode_relaxation_b} and \ref{fig:mode_relaxation_d} GHD simulations of a single, coherently excited density mode are compared to the predicted relaxation from Eq. (\ref{eq:mode_damping}). 
The simulations are carried out both in the Tonks-Girardeau and quasi-condensate regime, whereby the variance in advection velocity can be obtained analytically via Eqs. (\ref{eq:variance_sound_TG}, \ref{eq:variance_sound_QC}).
As evident from the figure, in the regime of relatively small coherent excitations and low temperatures, the relaxation of the mode is well-described by Eq. (\ref{eq:mode_damping}), illustrating how GHD based off the thermodynamic Bethe ansatz implicitly accounts for thermal fluctuations in the local velocity.

%% file: main_arXiv.bbl
\providecommand{\noopsort}[1]{}\providecommand{\singleletter}[1]{#1}%
\begin{thebibliography}{44}%
\makeatletter
\providecommand \@ifxundefined [1]{%
 \@ifx{#1\undefined}
}%
\providecommand \@ifnum [1]{%
 \ifnum #1\expandafter \@firstoftwo
 \else \expandafter \@secondoftwo
 \fi
}%
\providecommand \@ifx [1]{%
 \ifx #1\expandafter \@firstoftwo
 \else \expandafter \@secondoftwo
 \fi
}%
\providecommand \natexlab [1]{#1}%
\providecommand \enquote  [1]{``#1''}%
\providecommand \bibnamefont  [1]{#1}%
\providecommand \bibfnamefont [1]{#1}%
\providecommand \citenamefont [1]{#1}%
\providecommand \href@noop [0]{\@secondoftwo}%
\providecommand \href [0]{\begingroup \@sanitize@url \@href}%
\providecommand \@href[1]{\@@startlink{#1}\@@href}%
\providecommand \@@href[1]{\endgroup#1\@@endlink}%
\providecommand \@sanitize@url [0]{\catcode `\\12\catcode `\$12\catcode
  `\&12\catcode `\#12\catcode `\^12\catcode `\_12\catcode `\%12\relax}%
\providecommand \@@startlink[1]{}%
\providecommand \@@endlink[0]{}%
\providecommand \url  [0]{\begingroup\@sanitize@url \@url }%
\providecommand \@url [1]{\endgroup\@href {#1}{\urlprefix }}%
\providecommand \urlprefix  [0]{URL }%
\providecommand \Eprint [0]{\href }%
\providecommand \doibase [0]{https://doi.org/}%
\providecommand \selectlanguage [0]{\@gobble}%
\providecommand \bibinfo  [0]{\@secondoftwo}%
\providecommand \bibfield  [0]{\@secondoftwo}%
\providecommand \translation [1]{[#1]}%
\providecommand \BibitemOpen [0]{}%
\providecommand \bibitemStop [0]{}%
\providecommand \bibitemNoStop [0]{.\EOS\space}%
\providecommand \EOS [0]{\spacefactor3000\relax}%
\providecommand \BibitemShut  [1]{\csname bibitem#1\endcsname}%
\let\auto@bib@innerbib\@empty
\bibitem [{\citenamefont {Kr\"uger}\ \emph {et~al.}(2010)\citenamefont
  {Kr\"uger}, \citenamefont {Hofferberth}, \citenamefont {Mazets},
  \citenamefont {Lesanovsky},\ and\ \citenamefont
  {Schmiedmayer}}]{PhysRevLett.105.265302}%
  \BibitemOpen
  \bibfield  {author} {\bibinfo {author} {\bibfnamefont {P.}~\bibnamefont
  {Kr\"uger}}, \bibinfo {author} {\bibfnamefont {S.}~\bibnamefont
  {Hofferberth}}, \bibinfo {author} {\bibfnamefont {I.~E.}\ \bibnamefont
  {Mazets}}, \bibinfo {author} {\bibfnamefont {I.}~\bibnamefont {Lesanovsky}},\
  and\ \bibinfo {author} {\bibfnamefont {J.}~\bibnamefont {Schmiedmayer}},\
  }\bibfield  {title} {\bibinfo {title} {Weakly interacting {B}ose gas in the
  one-dimensional limit},\ }\href
  {https://doi.org/10.1103/PhysRevLett.105.265302} {\bibfield  {journal}
  {\bibinfo  {journal} {Phys. Rev. Lett.}\ }\textbf {\bibinfo {volume} {105}},\
  \bibinfo {pages} {265302} (\bibinfo {year} {2010})}\BibitemShut {NoStop}%
\bibitem [{\citenamefont {G\"orlitz}\ \emph {et~al.}(2001)\citenamefont
  {G\"orlitz}, \citenamefont {Vogels}, \citenamefont {Leanhardt}, \citenamefont
  {Raman}, \citenamefont {Gustavson}, \citenamefont {Abo-Shaeer}, \citenamefont
  {Chikkatur}, \citenamefont {Gupta}, \citenamefont {Inouye}, \citenamefont
  {Rosenband},\ and\ \citenamefont {Ketterle}}]{PhysRevLett.87.130402}%
  \BibitemOpen
  \bibfield  {author} {\bibinfo {author} {\bibfnamefont {A.}~\bibnamefont
  {G\"orlitz}}, \bibinfo {author} {\bibfnamefont {J.~M.}\ \bibnamefont
  {Vogels}}, \bibinfo {author} {\bibfnamefont {A.~E.}\ \bibnamefont
  {Leanhardt}}, \bibinfo {author} {\bibfnamefont {C.}~\bibnamefont {Raman}},
  \bibinfo {author} {\bibfnamefont {T.~L.}\ \bibnamefont {Gustavson}}, \bibinfo
  {author} {\bibfnamefont {J.~R.}\ \bibnamefont {Abo-Shaeer}}, \bibinfo
  {author} {\bibfnamefont {A.~P.}\ \bibnamefont {Chikkatur}}, \bibinfo {author}
  {\bibfnamefont {S.}~\bibnamefont {Gupta}}, \bibinfo {author} {\bibfnamefont
  {S.}~\bibnamefont {Inouye}}, \bibinfo {author} {\bibfnamefont
  {T.}~\bibnamefont {Rosenband}},\ and\ \bibinfo {author} {\bibfnamefont
  {W.}~\bibnamefont {Ketterle}},\ }\bibfield  {title} {\bibinfo {title}
  {Realization of {B}ose-{E}instein condensates in lower dimensions},\ }\href
  {https://doi.org/10.1103/PhysRevLett.87.130402} {\bibfield  {journal}
  {\bibinfo  {journal} {Phys. Rev. Lett.}\ }\textbf {\bibinfo {volume} {87}},\
  \bibinfo {pages} {130402} (\bibinfo {year} {2001})}\BibitemShut {NoStop}%
\bibitem [{\citenamefont {Kinoshita}\ \emph {et~al.}(2004)\citenamefont
  {Kinoshita}, \citenamefont {Wenger},\ and\ \citenamefont
  {Weiss}}]{doi:10.1126/science.1100700}%
  \BibitemOpen
  \bibfield  {author} {\bibinfo {author} {\bibfnamefont {T.}~\bibnamefont
  {Kinoshita}}, \bibinfo {author} {\bibfnamefont {T.}~\bibnamefont {Wenger}},\
  and\ \bibinfo {author} {\bibfnamefont {D.~S.}\ \bibnamefont {Weiss}},\
  }\bibfield  {title} {\bibinfo {title} {Observation of a one-dimensional
  {T}onks-{G}irardeau gas},\ }\href {https://doi.org/10.1126/science.1100700}
  {\bibfield  {journal} {\bibinfo  {journal} {Science}\ }\textbf {\bibinfo
  {volume} {305}},\ \bibinfo {pages} {1125} (\bibinfo {year}
  {2004})}\BibitemShut {NoStop}%
\bibitem [{\citenamefont {Bloch}\ \emph {et~al.}(2008)\citenamefont {Bloch},
  \citenamefont {Dalibard},\ and\ \citenamefont {Zwerger}}]{RevModPhys.80.885}%
  \BibitemOpen
  \bibfield  {author} {\bibinfo {author} {\bibfnamefont {I.}~\bibnamefont
  {Bloch}}, \bibinfo {author} {\bibfnamefont {J.}~\bibnamefont {Dalibard}},\
  and\ \bibinfo {author} {\bibfnamefont {W.}~\bibnamefont {Zwerger}},\
  }\bibfield  {title} {\bibinfo {title} {Many-body physics with ultracold
  gases},\ }\href {https://doi.org/10.1103/RevModPhys.80.885} {\bibfield
  {journal} {\bibinfo  {journal} {Rev. Mod. Phys.}\ }\textbf {\bibinfo {volume}
  {80}},\ \bibinfo {pages} {885} (\bibinfo {year} {2008})}\BibitemShut
  {NoStop}%
\bibitem [{\citenamefont {Dalfovo}\ \emph {et~al.}(1999)\citenamefont
  {Dalfovo}, \citenamefont {Giorgini}, \citenamefont {Pitaevskii},\ and\
  \citenamefont {Stringari}}]{DalfovoRev}%
  \BibitemOpen
  \bibfield  {author} {\bibinfo {author} {\bibfnamefont {F.}~\bibnamefont
  {Dalfovo}}, \bibinfo {author} {\bibfnamefont {S.}~\bibnamefont {Giorgini}},
  \bibinfo {author} {\bibfnamefont {L.~P.}\ \bibnamefont {Pitaevskii}},\ and\
  \bibinfo {author} {\bibfnamefont {S.}~\bibnamefont {Stringari}},\ }\bibfield
  {title} {\bibinfo {title} {Theory of {B}ose-{E}instein condensation in
  trapped gases},\ }\href {https://doi.org/10.1103/RevModPhys.71.463}
  {\bibfield  {journal} {\bibinfo  {journal} {Rev. Mod. Phys.}\ }\textbf
  {\bibinfo {volume} {71}},\ \bibinfo {pages} {463} (\bibinfo {year}
  {1999})}\BibitemShut {NoStop}%
\bibitem [{\citenamefont {Popov}\ and\ \citenamefont {Popov}(1987)}]{Popov87}%
  \BibitemOpen
  \bibfield  {author} {\bibinfo {author} {\bibfnamefont {V.~N.}\ \bibnamefont
  {Popov}}\ and\ \bibinfo {author} {\bibfnamefont {V.~N.}\ \bibnamefont
  {Popov}},\ }\href@noop {} {\emph {\bibinfo {title} {Functional integrals and
  collective excitations}}}\ (\bibinfo  {publisher} {Cambridge University
  Press},\ \bibinfo {year} {1987})\BibitemShut {NoStop}%
\bibitem [{\citenamefont {Esteve}\ \emph {et~al.}(2006)\citenamefont {Esteve},
  \citenamefont {Trebbia}, \citenamefont {Schumm}, \citenamefont {Aspect},
  \citenamefont {Westbrook},\ and\ \citenamefont
  {Bouchoule}}]{PhysRevLett.96.130403}%
  \BibitemOpen
  \bibfield  {author} {\bibinfo {author} {\bibfnamefont {J.}~\bibnamefont
  {Esteve}}, \bibinfo {author} {\bibfnamefont {J.-B.}\ \bibnamefont {Trebbia}},
  \bibinfo {author} {\bibfnamefont {T.}~\bibnamefont {Schumm}}, \bibinfo
  {author} {\bibfnamefont {A.}~\bibnamefont {Aspect}}, \bibinfo {author}
  {\bibfnamefont {C.~I.}\ \bibnamefont {Westbrook}},\ and\ \bibinfo {author}
  {\bibfnamefont {I.}~\bibnamefont {Bouchoule}},\ }\bibfield  {title} {\bibinfo
  {title} {Observations of density fluctuations in an elongated {B}ose gas:
  Ideal gas and quasicondensate regimes},\ }\href
  {https://doi.org/10.1103/PhysRevLett.96.130403} {\bibfield  {journal}
  {\bibinfo  {journal} {Phys. Rev. Lett.}\ }\textbf {\bibinfo {volume} {96}},\
  \bibinfo {pages} {130403} (\bibinfo {year} {2006})}\BibitemShut {NoStop}%
\bibitem [{\citenamefont {Armijo}\ \emph {et~al.}(2010)\citenamefont {Armijo},
  \citenamefont {Jacqmin}, \citenamefont {Kheruntsyan},\ and\ \citenamefont
  {Bouchoule}}]{PhysRevLett.105.230402}%
  \BibitemOpen
  \bibfield  {author} {\bibinfo {author} {\bibfnamefont {J.}~\bibnamefont
  {Armijo}}, \bibinfo {author} {\bibfnamefont {T.}~\bibnamefont {Jacqmin}},
  \bibinfo {author} {\bibfnamefont {K.~V.}\ \bibnamefont {Kheruntsyan}},\ and\
  \bibinfo {author} {\bibfnamefont {I.}~\bibnamefont {Bouchoule}},\ }\bibfield
  {title} {\bibinfo {title} {Probing three-body correlations in a quantum gas
  using the measurement of the third moment of density fluctuations},\ }\href
  {https://doi.org/10.1103/PhysRevLett.105.230402} {\bibfield  {journal}
  {\bibinfo  {journal} {Phys. Rev. Lett.}\ }\textbf {\bibinfo {volume} {105}},\
  \bibinfo {pages} {230402} (\bibinfo {year} {2010})}\BibitemShut {NoStop}%
\bibitem [{\citenamefont {Jacqmin}\ \emph {et~al.}(2011)\citenamefont
  {Jacqmin}, \citenamefont {Armijo}, \citenamefont {Berrada}, \citenamefont
  {Kheruntsyan},\ and\ \citenamefont {Bouchoule}}]{PhysRevLett.106.230405}%
  \BibitemOpen
  \bibfield  {author} {\bibinfo {author} {\bibfnamefont {T.}~\bibnamefont
  {Jacqmin}}, \bibinfo {author} {\bibfnamefont {J.}~\bibnamefont {Armijo}},
  \bibinfo {author} {\bibfnamefont {T.}~\bibnamefont {Berrada}}, \bibinfo
  {author} {\bibfnamefont {K.~V.}\ \bibnamefont {Kheruntsyan}},\ and\ \bibinfo
  {author} {\bibfnamefont {I.}~\bibnamefont {Bouchoule}},\ }\bibfield  {title}
  {\bibinfo {title} {Sub-poissonian fluctuations in a 1d {B}ose gas: From the
  quantum quasicondensate to the strongly interacting regime},\ }\href
  {https://doi.org/10.1103/PhysRevLett.106.230405} {\bibfield  {journal}
  {\bibinfo  {journal} {Phys. Rev. Lett.}\ }\textbf {\bibinfo {volume} {106}},\
  \bibinfo {pages} {230405} (\bibinfo {year} {2011})}\BibitemShut {NoStop}%
\bibitem [{\citenamefont {Imambekov}\ \emph {et~al.}(2009)\citenamefont
  {Imambekov}, \citenamefont {Mazets}, \citenamefont {Petrov}, \citenamefont
  {Gritsev}, \citenamefont {Manz}, \citenamefont {Hofferberth}, \citenamefont
  {Schumm}, \citenamefont {Demler},\ and\ \citenamefont
  {Schmiedmayer}}]{PhysRevA.80.033604}%
  \BibitemOpen
  \bibfield  {author} {\bibinfo {author} {\bibfnamefont {A.}~\bibnamefont
  {Imambekov}}, \bibinfo {author} {\bibfnamefont {I.~E.}\ \bibnamefont
  {Mazets}}, \bibinfo {author} {\bibfnamefont {D.~S.}\ \bibnamefont {Petrov}},
  \bibinfo {author} {\bibfnamefont {V.}~\bibnamefont {Gritsev}}, \bibinfo
  {author} {\bibfnamefont {S.}~\bibnamefont {Manz}}, \bibinfo {author}
  {\bibfnamefont {S.}~\bibnamefont {Hofferberth}}, \bibinfo {author}
  {\bibfnamefont {T.}~\bibnamefont {Schumm}}, \bibinfo {author} {\bibfnamefont
  {E.}~\bibnamefont {Demler}},\ and\ \bibinfo {author} {\bibfnamefont
  {J.}~\bibnamefont {Schmiedmayer}},\ }\bibfield  {title} {\bibinfo {title}
  {Density ripples in expanding low-dimensional gases as a probe of
  correlations},\ }\href {https://doi.org/10.1103/PhysRevA.80.033604}
  {\bibfield  {journal} {\bibinfo  {journal} {Phys. Rev. A}\ }\textbf {\bibinfo
  {volume} {80}},\ \bibinfo {pages} {033604} (\bibinfo {year}
  {2009})}\BibitemShut {NoStop}%
\bibitem [{\citenamefont {Schemmer}\ \emph {et~al.}(2018)\citenamefont
  {Schemmer}, \citenamefont {Johnson},\ and\ \citenamefont
  {Bouchoule}}]{PhysRevA.98.043604}%
  \BibitemOpen
  \bibfield  {author} {\bibinfo {author} {\bibfnamefont {M.}~\bibnamefont
  {Schemmer}}, \bibinfo {author} {\bibfnamefont {A.}~\bibnamefont {Johnson}},\
  and\ \bibinfo {author} {\bibfnamefont {I.}~\bibnamefont {Bouchoule}},\
  }\bibfield  {title} {\bibinfo {title} {Monitoring squeezed collective modes
  of a one-dimensional {B}ose gas after an interaction quench using
  density-ripple analysis},\ }\href
  {https://doi.org/10.1103/PhysRevA.98.043604} {\bibfield  {journal} {\bibinfo
  {journal} {Phys. Rev. A}\ }\textbf {\bibinfo {volume} {98}},\ \bibinfo
  {pages} {043604} (\bibinfo {year} {2018})}\BibitemShut {NoStop}%
\bibitem [{\citenamefont {Manz}\ \emph {et~al.}(2010)\citenamefont {Manz},
  \citenamefont {B\"ucker}, \citenamefont {Betz}, \citenamefont {Koller},
  \citenamefont {Hofferberth}, \citenamefont {Mazets}, \citenamefont
  {Imambekov}, \citenamefont {Demler}, \citenamefont {Perrin}, \citenamefont
  {Schmiedmayer},\ and\ \citenamefont {Schumm}}]{Manz10}%
  \BibitemOpen
  \bibfield  {author} {\bibinfo {author} {\bibfnamefont {S.}~\bibnamefont
  {Manz}}, \bibinfo {author} {\bibfnamefont {R.}~\bibnamefont {B\"ucker}},
  \bibinfo {author} {\bibfnamefont {T.}~\bibnamefont {Betz}}, \bibinfo {author}
  {\bibfnamefont {C.}~\bibnamefont {Koller}}, \bibinfo {author} {\bibfnamefont
  {S.}~\bibnamefont {Hofferberth}}, \bibinfo {author} {\bibfnamefont {I.~E.}\
  \bibnamefont {Mazets}}, \bibinfo {author} {\bibfnamefont {A.}~\bibnamefont
  {Imambekov}}, \bibinfo {author} {\bibfnamefont {E.}~\bibnamefont {Demler}},
  \bibinfo {author} {\bibfnamefont {A.}~\bibnamefont {Perrin}}, \bibinfo
  {author} {\bibfnamefont {J.}~\bibnamefont {Schmiedmayer}},\ and\ \bibinfo
  {author} {\bibfnamefont {T.}~\bibnamefont {Schumm}},\ }\bibfield  {title}
  {\bibinfo {title} {Two-point density correlations of quasicondensates in free
  expansion},\ }\href {https://doi.org/10.1103/PhysRevA.81.031610} {\bibfield
  {journal} {\bibinfo  {journal} {Phys. Rev. A}\ }\textbf {\bibinfo {volume}
  {81}},\ \bibinfo {pages} {031610(R)} (\bibinfo {year} {2010})}\BibitemShut
  {NoStop}%
\bibitem [{\citenamefont {M\o{}ller}\ \emph
  {et~al.}(2021{\natexlab{a}})\citenamefont {M\o{}ller}, \citenamefont
  {Schweigler}, \citenamefont {Tajik}, \citenamefont {Sabino}, \citenamefont
  {Cataldini}, \citenamefont {Ji},\ and\ \citenamefont
  {Schmiedmayer}}]{PhysRevA.104.043305}%
  \BibitemOpen
  \bibfield  {author} {\bibinfo {author} {\bibfnamefont {F.}~\bibnamefont
  {M\o{}ller}}, \bibinfo {author} {\bibfnamefont {T.}~\bibnamefont
  {Schweigler}}, \bibinfo {author} {\bibfnamefont {M.}~\bibnamefont {Tajik}},
  \bibinfo {author} {\bibfnamefont {J.}~\bibnamefont {Sabino}}, \bibinfo
  {author} {\bibfnamefont {F.}~\bibnamefont {Cataldini}}, \bibinfo {author}
  {\bibfnamefont {S.-C.}\ \bibnamefont {Ji}},\ and\ \bibinfo {author}
  {\bibfnamefont {J.}~\bibnamefont {Schmiedmayer}},\ }\bibfield  {title}
  {\bibinfo {title} {Thermometry of one-dimensional {B}ose gases with neural
  networks},\ }\href {https://doi.org/10.1103/PhysRevA.104.043305} {\bibfield
  {journal} {\bibinfo  {journal} {Phys. Rev. A}\ }\textbf {\bibinfo {volume}
  {104}},\ \bibinfo {pages} {043305} (\bibinfo {year}
  {2021}{\natexlab{a}})}\BibitemShut {NoStop}%
\bibitem [{\citenamefont {Mora}\ and\ \citenamefont
  {Castin}(2003)}]{PhysRevA.67.053615}%
  \BibitemOpen
  \bibfield  {author} {\bibinfo {author} {\bibfnamefont {C.}~\bibnamefont
  {Mora}}\ and\ \bibinfo {author} {\bibfnamefont {Y.}~\bibnamefont {Castin}},\
  }\bibfield  {title} {\bibinfo {title} {Extension of {B}ogoliubov theory to
  quasicondensates},\ }\href {https://doi.org/10.1103/PhysRevA.67.053615}
  {\bibfield  {journal} {\bibinfo  {journal} {Phys. Rev. A}\ }\textbf {\bibinfo
  {volume} {67}},\ \bibinfo {pages} {053615} (\bibinfo {year}
  {2003})}\BibitemShut {NoStop}%
\bibitem [{\citenamefont {Haldane}(1981)}]{Haldane81}%
  \BibitemOpen
  \bibfield  {author} {\bibinfo {author} {\bibfnamefont {F.~D.~M.}\
  \bibnamefont {Haldane}},\ }\bibfield  {title} {\bibinfo {title}
  {{\textquotesingle}{L}uttinger liquid theory{\textquotesingle} of
  one-dimensional quantum fluids. {I}. properties of the {L}uttinger model and
  their extension to the general 1d interacting spinless {F}ermi gas},\ }\href
  {https://doi.org/10.1088/0022-3719/14/19/010} {\bibfield  {journal} {\bibinfo
   {journal} {Journal of Physics C: Solid State Physics}\ }\textbf {\bibinfo
  {volume} {14}},\ \bibinfo {pages} {2585} (\bibinfo {year}
  {1981})}\BibitemShut {NoStop}%
\bibitem [{\citenamefont {Haldane}(1994)}]{Haldane94}%
  \BibitemOpen
  \bibfield  {author} {\bibinfo {author} {\bibfnamefont {F.~D.~M.}\
  \bibnamefont {Haldane}},\ }\bibfield  {title} {\bibinfo {title}
  {{L}uttinger's theorem and bosonization of the {F}ermi surface},\ }in\
  \href@noop {} {\emph {\bibinfo {booktitle} {Proceedings of the International
  School of Physics "Enrico Fermi", Course CXXI "Perspectives in Many-Particle
  Physics"}}},\ \bibinfo {editor} {edited by\ \bibinfo {editor} {\bibfnamefont
  {R.~A.}\ \bibnamefont {Broglia}}\ and\ \bibinfo {editor} {\bibfnamefont
  {J.~R.}\ \bibnamefont {Schrieffer}}}\ (\bibinfo {year} {1994})\ pp.\ \bibinfo
  {pages} {5--29}\BibitemShut {NoStop}%
\bibitem [{\citenamefont {Cazalilla}(2004)}]{Cazalilla04}%
  \BibitemOpen
  \bibfield  {author} {\bibinfo {author} {\bibfnamefont {M.~A.}\ \bibnamefont
  {Cazalilla}},\ }\bibfield  {title} {\bibinfo {title} {Bosonizing
  one-dimensional cold atomic gases},\ }\href
  {https://doi.org/10.1088/0953-4075/37/7/051} {\bibfield  {journal} {\bibinfo
  {journal} {Journal of Physics B: Atomic, Molecular and Optical Physics}\
  }\textbf {\bibinfo {volume} {37}},\ \bibinfo {pages} {S1} (\bibinfo {year}
  {2004})}\BibitemShut {NoStop}%
\bibitem [{\citenamefont {Leggett}(2001)}]{RevModPhys.73.307}%
  \BibitemOpen
  \bibfield  {author} {\bibinfo {author} {\bibfnamefont {A.~J.}\ \bibnamefont
  {Leggett}},\ }\bibfield  {title} {\bibinfo {title} {{B}ose-{E}instein
  condensation in the alkali gases: Some fundamental concepts},\ }\href
  {https://doi.org/10.1103/RevModPhys.73.307} {\bibfield  {journal} {\bibinfo
  {journal} {Rev. Mod. Phys.}\ }\textbf {\bibinfo {volume} {73}},\ \bibinfo
  {pages} {307} (\bibinfo {year} {2001})}\BibitemShut {NoStop}%
\bibitem [{\citenamefont {Castro-Alvaredo}\ \emph {et~al.}(2016)\citenamefont
  {Castro-Alvaredo}, \citenamefont {Doyon},\ and\ \citenamefont
  {Yoshimura}}]{castro2016emergent}%
  \BibitemOpen
  \bibfield  {author} {\bibinfo {author} {\bibfnamefont {O.~A.}\ \bibnamefont
  {Castro-Alvaredo}}, \bibinfo {author} {\bibfnamefont {B.}~\bibnamefont
  {Doyon}},\ and\ \bibinfo {author} {\bibfnamefont {T.}~\bibnamefont
  {Yoshimura}},\ }\bibfield  {title} {\bibinfo {title} {Emergent hydrodynamics
  in integrable quantum systems out of equilibrium},\ }\href
  {https://doi.org/10.1103/PhysRevX.6.041065} {\bibfield  {journal} {\bibinfo
  {journal} {\textit{Phys. Rev. X}}\ }\textbf {\bibinfo {volume} {6}},\
  \bibinfo {pages} {041065} (\bibinfo {year} {2016})}\BibitemShut {NoStop}%
\bibitem [{\citenamefont {Bertini}\ \emph {et~al.}(2016)\citenamefont
  {Bertini}, \citenamefont {Collura}, \citenamefont {De~Nardis},\ and\
  \citenamefont {Fagotti}}]{bertini2016transport}%
  \BibitemOpen
  \bibfield  {author} {\bibinfo {author} {\bibfnamefont {B.}~\bibnamefont
  {Bertini}}, \bibinfo {author} {\bibfnamefont {M.}~\bibnamefont {Collura}},
  \bibinfo {author} {\bibfnamefont {J.}~\bibnamefont {De~Nardis}},\ and\
  \bibinfo {author} {\bibfnamefont {M.}~\bibnamefont {Fagotti}},\ }\bibfield
  {title} {\bibinfo {title} {Transport in out-of-equilibrium {XXZ} chains:
  Exact profiles of charges and currents},\ }\href
  {https://doi.org/10.1103/PhysRevLett.117.207201} {\bibfield  {journal}
  {\bibinfo  {journal} {\textit{Phys. Rev. Lett.}}\ }\textbf {\bibinfo {volume}
  {117}},\ \bibinfo {pages} {207201} (\bibinfo {year} {2016})}\BibitemShut
  {NoStop}%
\bibitem [{\citenamefont {Lieb}\ and\ \citenamefont {Liniger}(1963)}]{LL1}%
  \BibitemOpen
  \bibfield  {author} {\bibinfo {author} {\bibfnamefont {E.~H.}\ \bibnamefont
  {Lieb}}\ and\ \bibinfo {author} {\bibfnamefont {W.}~\bibnamefont {Liniger}},\
  }\bibfield  {title} {\bibinfo {title} {Exact analysis of an interacting
  {B}ose gas. {I}. the general solution and the ground state},\ }\href
  {https://doi.org/10.1103/PhysRev.130.1605} {\bibfield  {journal} {\bibinfo
  {journal} {Phys. Rev.}\ }\textbf {\bibinfo {volume} {130}},\ \bibinfo {pages}
  {1605} (\bibinfo {year} {1963})}\BibitemShut {NoStop}%
\bibitem [{\citenamefont {Lieb}(1963)}]{LL2}%
  \BibitemOpen
  \bibfield  {author} {\bibinfo {author} {\bibfnamefont {E.~H.}\ \bibnamefont
  {Lieb}},\ }\bibfield  {title} {\bibinfo {title} {Exact analysis of an
  interacting {B}ose gas. {II}. the excitation spectrum},\ }\href
  {https://doi.org/10.1103/PhysRev.130.1616} {\bibfield  {journal} {\bibinfo
  {journal} {Phys. Rev.}\ }\textbf {\bibinfo {volume} {130}},\ \bibinfo {pages}
  {1616} (\bibinfo {year} {1963})}\BibitemShut {NoStop}%
\bibitem [{\citenamefont {Yang}\ and\ \citenamefont {Yang}(1969)}]{YY1}%
  \BibitemOpen
  \bibfield  {author} {\bibinfo {author} {\bibfnamefont {C.~N.}\ \bibnamefont
  {Yang}}\ and\ \bibinfo {author} {\bibfnamefont {C.~P.}\ \bibnamefont
  {Yang}},\ }\bibfield  {title} {\bibinfo {title} {Thermodynamics of a
  one‐dimensional system of bosons with repulsive delta‐function
  interaction},\ }\href {https://doi.org/10.1063/1.1664947} {\bibfield
  {journal} {\bibinfo  {journal} {Journal of Mathematical Physics}\ }\textbf
  {\bibinfo {volume} {10}},\ \bibinfo {pages} {1115} (\bibinfo {year}
  {1969})}\BibitemShut {NoStop}%
\bibitem [{\citenamefont {Schemmer}\ \emph {et~al.}(2019)\citenamefont
  {Schemmer}, \citenamefont {Bouchoule}, \citenamefont {Doyon},\ and\
  \citenamefont {Dubail}}]{schemmer2019generalized}%
  \BibitemOpen
  \bibfield  {author} {\bibinfo {author} {\bibfnamefont {M.}~\bibnamefont
  {Schemmer}}, \bibinfo {author} {\bibfnamefont {I.}~\bibnamefont {Bouchoule}},
  \bibinfo {author} {\bibfnamefont {B.}~\bibnamefont {Doyon}},\ and\ \bibinfo
  {author} {\bibfnamefont {J.}~\bibnamefont {Dubail}},\ }\bibfield  {title}
  {\bibinfo {title} {Generalized {H}ydrodynamics on an atom chip},\ }\href
  {https://doi.org/10.1103/PhysRevLett.122.090601} {\bibfield  {journal}
  {\bibinfo  {journal} {Phys. Rev. Lett.}\ }\textbf {\bibinfo {volume} {122}},\
  \bibinfo {pages} {090601} (\bibinfo {year} {2019})}\BibitemShut {NoStop}%
\bibitem [{\citenamefont {M\o{}ller}\ \emph
  {et~al.}(2021{\natexlab{b}})\citenamefont {M\o{}ller}, \citenamefont {Li},
  \citenamefont {Mazets}, \citenamefont {Stimming}, \citenamefont {Zhou},
  \citenamefont {Zhu}, \citenamefont {Chen},\ and\ \citenamefont
  {Schmiedmayer}}]{PhysRevLett.126.090602}%
  \BibitemOpen
  \bibfield  {author} {\bibinfo {author} {\bibfnamefont {F.}~\bibnamefont
  {M\o{}ller}}, \bibinfo {author} {\bibfnamefont {C.}~\bibnamefont {Li}},
  \bibinfo {author} {\bibfnamefont {I.}~\bibnamefont {Mazets}}, \bibinfo
  {author} {\bibfnamefont {H.-P.}\ \bibnamefont {Stimming}}, \bibinfo {author}
  {\bibfnamefont {T.}~\bibnamefont {Zhou}}, \bibinfo {author} {\bibfnamefont
  {Z.}~\bibnamefont {Zhu}}, \bibinfo {author} {\bibfnamefont {X.}~\bibnamefont
  {Chen}},\ and\ \bibinfo {author} {\bibfnamefont {J.}~\bibnamefont
  {Schmiedmayer}},\ }\bibfield  {title} {\bibinfo {title} {Extension of the
  generalized hydrodynamics to the dimensional crossover regime},\ }\href
  {https://doi.org/10.1103/PhysRevLett.126.090602} {\bibfield  {journal}
  {\bibinfo  {journal} {Phys. Rev. Lett.}\ }\textbf {\bibinfo {volume} {126}},\
  \bibinfo {pages} {090602} (\bibinfo {year} {2021}{\natexlab{b}})}\BibitemShut
  {NoStop}%
\bibitem [{\citenamefont {Malvania}\ \emph {et~al.}(2021)\citenamefont
  {Malvania}, \citenamefont {Zhang}, \citenamefont {Le}, \citenamefont
  {Dubail}, \citenamefont {Rigol},\ and\ \citenamefont
  {Weiss}}]{malvania2020generalized}%
  \BibitemOpen
  \bibfield  {author} {\bibinfo {author} {\bibfnamefont {N.}~\bibnamefont
  {Malvania}}, \bibinfo {author} {\bibfnamefont {Y.}~\bibnamefont {Zhang}},
  \bibinfo {author} {\bibfnamefont {Y.}~\bibnamefont {Le}}, \bibinfo {author}
  {\bibfnamefont {J.}~\bibnamefont {Dubail}}, \bibinfo {author} {\bibfnamefont
  {M.}~\bibnamefont {Rigol}},\ and\ \bibinfo {author} {\bibfnamefont {D.~S.}\
  \bibnamefont {Weiss}},\ }\bibfield  {title} {\bibinfo {title} {Generalized
  hydrodynamics in strongly interacting 1d {B}ose gases},\ }\href
  {https://doi.org/10.1126/science.abf0147} {\bibfield  {journal} {\bibinfo
  {journal} {Science}\ }\textbf {\bibinfo {volume} {373}},\ \bibinfo {pages}
  {1129} (\bibinfo {year} {2021})}\BibitemShut {NoStop}%
\bibitem [{\citenamefont {Cataldini}\ \emph {et~al.}(2021)\citenamefont
  {Cataldini}, \citenamefont {M{\o}ller}, \citenamefont {Tajik}, \citenamefont
  {Sabino}, \citenamefont {Schweigler}, \citenamefont {Ji}, \citenamefont
  {Rauer},\ and\ \citenamefont {Schmiedmayer}}]{cataldini2021emergent}%
  \BibitemOpen
  \bibfield  {author} {\bibinfo {author} {\bibfnamefont {F.}~\bibnamefont
  {Cataldini}}, \bibinfo {author} {\bibfnamefont {F.}~\bibnamefont
  {M{\o}ller}}, \bibinfo {author} {\bibfnamefont {M.}~\bibnamefont {Tajik}},
  \bibinfo {author} {\bibfnamefont {J.}~\bibnamefont {Sabino}}, \bibinfo
  {author} {\bibfnamefont {T.}~\bibnamefont {Schweigler}}, \bibinfo {author}
  {\bibfnamefont {S.-C.}\ \bibnamefont {Ji}}, \bibinfo {author} {\bibfnamefont
  {B.}~\bibnamefont {Rauer}},\ and\ \bibinfo {author} {\bibfnamefont
  {J.}~\bibnamefont {Schmiedmayer}},\ }\bibfield  {title} {\bibinfo {title}
  {Emergent {P}auli blocking in a weakly interacting {B}ose gas},\ }\bibfield
  {journal} {\bibinfo  {journal} {arXiv preprint arXiv:2111.13647}\ }\href
  {https://doi.org/10.48550/arXiv.2111.13647} {10.48550/arXiv.2111.13647}
  (\bibinfo {year} {2021})\BibitemShut {NoStop}%
\bibitem [{\citenamefont {Doyon}(2018)}]{doyon2018exact}%
  \BibitemOpen
  \bibfield  {author} {\bibinfo {author} {\bibfnamefont {B.}~\bibnamefont
  {Doyon}},\ }\bibfield  {title} {\bibinfo {title} {{Exact large-scale
  correlations in integrable systems out of equilibrium}},\ }\href
  {https://doi.org/10.21468/SciPostPhys.5.5.054} {\bibfield  {journal}
  {\bibinfo  {journal} {SciPost Phys.}\ }\textbf {\bibinfo {volume} {5}},\
  \bibinfo {pages} {54} (\bibinfo {year} {2018})}\BibitemShut {NoStop}%
\bibitem [{\citenamefont {Møller}\ \emph {et~al.}(2020)\citenamefont
  {Møller}, \citenamefont {Perfetto}, \citenamefont {Doyon},\ and\
  \citenamefont {Schmiedmayer}}]{10.21468/SciPostPhysCore.3.2.016}%
  \BibitemOpen
  \bibfield  {author} {\bibinfo {author} {\bibfnamefont {F.~S.}\ \bibnamefont
  {Møller}}, \bibinfo {author} {\bibfnamefont {G.}~\bibnamefont {Perfetto}},
  \bibinfo {author} {\bibfnamefont {B.}~\bibnamefont {Doyon}},\ and\ \bibinfo
  {author} {\bibfnamefont {J.}~\bibnamefont {Schmiedmayer}},\ }\bibfield
  {title} {\bibinfo {title} {{Euler-scale dynamical correlations in integrable
  systems with fluid motion}},\ }\href
  {https://doi.org/10.21468/SciPostPhysCore.3.2.016} {\bibfield  {journal}
  {\bibinfo  {journal} {SciPost Phys. Core}\ }\textbf {\bibinfo {volume} {3}},\
  \bibinfo {pages} {16} (\bibinfo {year} {2020})}\BibitemShut {NoStop}%
\bibitem [{\citenamefont {Hofferberth}\ \emph {et~al.}(2007)\citenamefont
  {Hofferberth}, \citenamefont {Lesanovsky}, \citenamefont {Fischer},
  \citenamefont {Schumm},\ and\ \citenamefont {Schmiedmayer}}]{Hofferberth07}%
  \BibitemOpen
  \bibfield  {author} {\bibinfo {author} {\bibfnamefont {S.}~\bibnamefont
  {Hofferberth}}, \bibinfo {author} {\bibfnamefont {I.}~\bibnamefont
  {Lesanovsky}}, \bibinfo {author} {\bibfnamefont {B.}~\bibnamefont {Fischer}},
  \bibinfo {author} {\bibfnamefont {T.}~\bibnamefont {Schumm}},\ and\ \bibinfo
  {author} {\bibfnamefont {J.}~\bibnamefont {Schmiedmayer}},\ }\bibfield
  {title} {\bibinfo {title} {Non-equilibrium coherence dynamics in
  one-dimensional {B}ose gases},\ }\href {https://doi.org/10.1038/nature06149}
  {\bibfield  {journal} {\bibinfo  {journal} {Nature}\ }\textbf {\bibinfo
  {volume} {449}},\ \bibinfo {pages} {324} (\bibinfo {year}
  {2007})}\BibitemShut {NoStop}%
\bibitem [{\citenamefont {Gring}\ \emph {et~al.}(2012)\citenamefont {Gring},
  \citenamefont {Kuhnert}, \citenamefont {Langen}, \citenamefont {Kitagawa},
  \citenamefont {Rauer}, \citenamefont {Schreitl}, \citenamefont {Mazets},
  \citenamefont {Smith}, \citenamefont {Demler},\ and\ \citenamefont
  {Schmiedmayer}}]{Gring1318}%
  \BibitemOpen
  \bibfield  {author} {\bibinfo {author} {\bibfnamefont {M.}~\bibnamefont
  {Gring}}, \bibinfo {author} {\bibfnamefont {M.}~\bibnamefont {Kuhnert}},
  \bibinfo {author} {\bibfnamefont {T.}~\bibnamefont {Langen}}, \bibinfo
  {author} {\bibfnamefont {T.}~\bibnamefont {Kitagawa}}, \bibinfo {author}
  {\bibfnamefont {B.}~\bibnamefont {Rauer}}, \bibinfo {author} {\bibfnamefont
  {M.}~\bibnamefont {Schreitl}}, \bibinfo {author} {\bibfnamefont
  {I.}~\bibnamefont {Mazets}}, \bibinfo {author} {\bibfnamefont {D.~A.}\
  \bibnamefont {Smith}}, \bibinfo {author} {\bibfnamefont {E.}~\bibnamefont
  {Demler}},\ and\ \bibinfo {author} {\bibfnamefont {J.}~\bibnamefont
  {Schmiedmayer}},\ }\bibfield  {title} {\bibinfo {title} {Relaxation and
  prethermalization in an isolated quantum system},\ }\href
  {https://doi.org/10.1126/science.1224953} {\bibfield  {journal} {\bibinfo
  {journal} {Science}\ }\textbf {\bibinfo {volume} {337}},\ \bibinfo {pages}
  {1318} (\bibinfo {year} {2012})}\BibitemShut {NoStop}%
\bibitem [{\citenamefont {Langen}\ \emph {et~al.}(2015)\citenamefont {Langen},
  \citenamefont {Erne}, \citenamefont {Geiger}, \citenamefont {Rauer},
  \citenamefont {Schweigler}, \citenamefont {Kuhnert}, \citenamefont
  {Rohringer}, \citenamefont {Mazets}, \citenamefont {Gasenzer},\ and\
  \citenamefont {Schmiedmayer}}]{doi:10.1126/science.1257026}%
  \BibitemOpen
  \bibfield  {author} {\bibinfo {author} {\bibfnamefont {T.}~\bibnamefont
  {Langen}}, \bibinfo {author} {\bibfnamefont {S.}~\bibnamefont {Erne}},
  \bibinfo {author} {\bibfnamefont {R.}~\bibnamefont {Geiger}}, \bibinfo
  {author} {\bibfnamefont {B.}~\bibnamefont {Rauer}}, \bibinfo {author}
  {\bibfnamefont {T.}~\bibnamefont {Schweigler}}, \bibinfo {author}
  {\bibfnamefont {M.}~\bibnamefont {Kuhnert}}, \bibinfo {author} {\bibfnamefont
  {W.}~\bibnamefont {Rohringer}}, \bibinfo {author} {\bibfnamefont {I.~E.}\
  \bibnamefont {Mazets}}, \bibinfo {author} {\bibfnamefont {T.}~\bibnamefont
  {Gasenzer}},\ and\ \bibinfo {author} {\bibfnamefont {J.}~\bibnamefont
  {Schmiedmayer}},\ }\bibfield  {title} {\bibinfo {title} {Experimental
  observation of a generalized {G}ibbs ensemble},\ }\href
  {https://doi.org/10.1126/science.1257026} {\bibfield  {journal} {\bibinfo
  {journal} {Science}\ }\textbf {\bibinfo {volume} {348}},\ \bibinfo {pages}
  {207} (\bibinfo {year} {2015})}\BibitemShut {NoStop}%
\bibitem [{\citenamefont {Schweigler}\ \emph {et~al.}(2017)\citenamefont
  {Schweigler}, \citenamefont {Kasper}, \citenamefont {Erne}, \citenamefont
  {Mazets}, \citenamefont {Rauer}, \citenamefont {Cataldini}, \citenamefont
  {Langen}, \citenamefont {Gasenzer}, \citenamefont {Berges},\ and\
  \citenamefont {Schmiedmayer}}]{Schweigler17}%
  \BibitemOpen
  \bibfield  {author} {\bibinfo {author} {\bibfnamefont {T.}~\bibnamefont
  {Schweigler}}, \bibinfo {author} {\bibfnamefont {V.}~\bibnamefont {Kasper}},
  \bibinfo {author} {\bibfnamefont {S.}~\bibnamefont {Erne}}, \bibinfo {author}
  {\bibfnamefont {I.}~\bibnamefont {Mazets}}, \bibinfo {author} {\bibfnamefont
  {B.}~\bibnamefont {Rauer}}, \bibinfo {author} {\bibfnamefont
  {F.}~\bibnamefont {Cataldini}}, \bibinfo {author} {\bibfnamefont
  {T.}~\bibnamefont {Langen}}, \bibinfo {author} {\bibfnamefont
  {T.}~\bibnamefont {Gasenzer}}, \bibinfo {author} {\bibfnamefont
  {J.}~\bibnamefont {Berges}},\ and\ \bibinfo {author} {\bibfnamefont
  {J.}~\bibnamefont {Schmiedmayer}},\ }\bibfield  {title} {\bibinfo {title}
  {Experimental characterization of a quantum many-body system via higher-order
  correlations},\ }\href {https://doi.org/10.1038/nature22310} {\bibfield
  {journal} {\bibinfo  {journal} {Nature}\ }\textbf {\bibinfo {volume} {545}},\
  \bibinfo {pages} {323} (\bibinfo {year} {2017})}\BibitemShut {NoStop}%
\bibitem [{\citenamefont {Rauer}\ \emph {et~al.}(2018)\citenamefont {Rauer},
  \citenamefont {Erne}, \citenamefont {Thomas}, \citenamefont {Cataldini},
  \citenamefont {Tajik},\ and\ \citenamefont
  {Schmiedmayer}}]{RauerRecurrencies}%
  \BibitemOpen
  \bibfield  {author} {\bibinfo {author} {\bibfnamefont {B.}~\bibnamefont
  {Rauer}}, \bibinfo {author} {\bibfnamefont {S.}~\bibnamefont {Erne}},
  \bibinfo {author} {\bibfnamefont {S.}~\bibnamefont {Thomas}}, \bibinfo
  {author} {\bibfnamefont {F.}~\bibnamefont {Cataldini}}, \bibinfo {author}
  {\bibfnamefont {M.}~\bibnamefont {Tajik}},\ and\ \bibinfo {author}
  {\bibfnamefont {J.}~\bibnamefont {Schmiedmayer}},\ }\bibfield  {title}
  {\bibinfo {title} {Recurrences in an isolated quantum many-body system},\
  }\href {https://doi.org/10.1126/science.aan7938} {\bibfield  {journal}
  {\bibinfo  {journal} {Science}\ }\textbf {\bibinfo {volume} {360}},\ \bibinfo
  {pages} {307} (\bibinfo {year} {2018})}\BibitemShut {NoStop}%
\bibitem [{\citenamefont {Schweigler}\ \emph {et~al.}(2021)\citenamefont
  {Schweigler}, \citenamefont {Gluza}, \citenamefont {Tajik}, \citenamefont
  {Sotiriadis}, \citenamefont {Cataldini}, \citenamefont {Ji}, \citenamefont
  {M{\o}ller}, \citenamefont {Sabino}, \citenamefont {Rauer}, \citenamefont
  {Eisert},\ and\ \citenamefont {Schmiedmayer}}]{Schweigler2021}%
  \BibitemOpen
  \bibfield  {author} {\bibinfo {author} {\bibfnamefont {T.}~\bibnamefont
  {Schweigler}}, \bibinfo {author} {\bibfnamefont {M.}~\bibnamefont {Gluza}},
  \bibinfo {author} {\bibfnamefont {M.}~\bibnamefont {Tajik}}, \bibinfo
  {author} {\bibfnamefont {S.}~\bibnamefont {Sotiriadis}}, \bibinfo {author}
  {\bibfnamefont {F.}~\bibnamefont {Cataldini}}, \bibinfo {author}
  {\bibfnamefont {S.-C.}\ \bibnamefont {Ji}}, \bibinfo {author} {\bibfnamefont
  {F.~S.}\ \bibnamefont {M{\o}ller}}, \bibinfo {author} {\bibfnamefont
  {J.}~\bibnamefont {Sabino}}, \bibinfo {author} {\bibfnamefont
  {B.}~\bibnamefont {Rauer}}, \bibinfo {author} {\bibfnamefont
  {J.}~\bibnamefont {Eisert}},\ and\ \bibinfo {author} {\bibfnamefont
  {J.}~\bibnamefont {Schmiedmayer}},\ }\bibfield  {title} {\bibinfo {title}
  {Decay and recurrence of non-{G}aussian correlations in a quantum many-body
  system},\ }\href {https://doi.org/10.1038/s41567-020-01139-2} {\bibfield
  {journal} {\bibinfo  {journal} {Nature Physics}\ }\textbf {\bibinfo {volume}
  {17}},\ \bibinfo {pages} {559} (\bibinfo {year} {2021})}\BibitemShut
  {NoStop}%
\bibitem [{Note1()}]{Note1}%
  \BibitemOpen
  \bibinfo {note} {See Supplemental Material for further details.}\BibitemShut
  {Stop}%
\bibitem [{\citenamefont {Doyon}\ \emph {et~al.}(2017)\citenamefont {Doyon},
  \citenamefont {Dubail}, \citenamefont {Konik},\ and\ \citenamefont
  {Yoshimura}}]{Doyon17a}%
  \BibitemOpen
  \bibfield  {author} {\bibinfo {author} {\bibfnamefont {B.}~\bibnamefont
  {Doyon}}, \bibinfo {author} {\bibfnamefont {J.}~\bibnamefont {Dubail}},
  \bibinfo {author} {\bibfnamefont {R.}~\bibnamefont {Konik}},\ and\ \bibinfo
  {author} {\bibfnamefont {T.}~\bibnamefont {Yoshimura}},\ }\bibfield  {title}
  {\bibinfo {title} {Large-scale description of interacting one-dimensional
  {B}ose gases: {G}eneralized {H}ydrodynamics supersedes conventional
  hydrodynamics},\ }\href {https://doi.org/10.1103/PhysRevLett.119.195301}
  {\bibfield  {journal} {\bibinfo  {journal} {Phys. Rev. Lett.}\ }\textbf
  {\bibinfo {volume} {119}},\ \bibinfo {pages} {195301} (\bibinfo {year}
  {2017})}\BibitemShut {NoStop}%
\bibitem [{\citenamefont {Fagotti}(2020)}]{10.21468/SciPostPhys.8.3.048}%
  \BibitemOpen
  \bibfield  {author} {\bibinfo {author} {\bibfnamefont {M.}~\bibnamefont
  {Fagotti}},\ }\bibfield  {title} {\bibinfo {title} {{Locally quasi-stationary
  states in noninteracting spin chains}},\ }\href
  {https://doi.org/10.21468/SciPostPhys.8.3.048} {\bibfield  {journal}
  {\bibinfo  {journal} {SciPost Phys.}\ }\textbf {\bibinfo {volume} {8}},\
  \bibinfo {pages} {48} (\bibinfo {year} {2020})}\BibitemShut {NoStop}%
\bibitem [{\citenamefont {Fagotti}(2017)}]{PhysRevB.96.220302}%
  \BibitemOpen
  \bibfield  {author} {\bibinfo {author} {\bibfnamefont {M.}~\bibnamefont
  {Fagotti}},\ }\bibfield  {title} {\bibinfo {title} {Higher-order generalized
  hydrodynamics in one dimension: The noninteracting test},\ }\href
  {https://doi.org/10.1103/PhysRevB.96.220302} {\bibfield  {journal} {\bibinfo
  {journal} {Phys. Rev. B}\ }\textbf {\bibinfo {volume} {96}},\ \bibinfo
  {pages} {220302(R)} (\bibinfo {year} {2017})}\BibitemShut {NoStop}%
\bibitem [{\citenamefont {Ruggiero}\ \emph {et~al.}(2020)\citenamefont
  {Ruggiero}, \citenamefont {Calabrese}, \citenamefont {Doyon},\ and\
  \citenamefont {Dubail}}]{PhysRevLett.124.140603}%
  \BibitemOpen
  \bibfield  {author} {\bibinfo {author} {\bibfnamefont {P.}~\bibnamefont
  {Ruggiero}}, \bibinfo {author} {\bibfnamefont {P.}~\bibnamefont {Calabrese}},
  \bibinfo {author} {\bibfnamefont {B.}~\bibnamefont {Doyon}},\ and\ \bibinfo
  {author} {\bibfnamefont {J.}~\bibnamefont {Dubail}},\ }\bibfield  {title}
  {\bibinfo {title} {Quantum generalized hydrodynamics},\ }\href
  {https://doi.org/10.1103/PhysRevLett.124.140603} {\bibfield  {journal}
  {\bibinfo  {journal} {Phys. Rev. Lett.}\ }\textbf {\bibinfo {volume} {124}},\
  \bibinfo {pages} {140603} (\bibinfo {year} {2020})}\BibitemShut {NoStop}%
\bibitem [{\citenamefont {Ruggiero}\ \emph {et~al.}(2021)\citenamefont
  {Ruggiero}, \citenamefont {Calabrese}, \citenamefont {Doyon},\ and\
  \citenamefont {Dubail}}]{Ruggiero_2021}%
  \BibitemOpen
  \bibfield  {author} {\bibinfo {author} {\bibfnamefont {P.}~\bibnamefont
  {Ruggiero}}, \bibinfo {author} {\bibfnamefont {P.}~\bibnamefont {Calabrese}},
  \bibinfo {author} {\bibfnamefont {B.}~\bibnamefont {Doyon}},\ and\ \bibinfo
  {author} {\bibfnamefont {J.}~\bibnamefont {Dubail}},\ }\bibfield  {title}
  {\bibinfo {title} {Quantum generalized hydrodynamics of the
  {T}onks{\textendash}{G}irardeau gas: density fluctuations and entanglement
  entropy},\ }\href {https://doi.org/10.1088/1751-8121/ac3d68} {\bibfield
  {journal} {\bibinfo  {journal} {Journal of Physics A: Mathematical and
  Theoretical}\ }\textbf {\bibinfo {volume} {55}},\ \bibinfo {pages} {024003}
  (\bibinfo {year} {2021})}\BibitemShut {NoStop}%
\bibitem [{\citenamefont {Hudson}(1964)}]{Hudson63}%
  \BibitemOpen
  \bibfield  {author} {\bibinfo {author} {\bibfnamefont {D.~J.}\ \bibnamefont
  {Hudson}},\ }\href@noop {} {\emph {\bibinfo {title} {Lectures on elementary
  statistics and probability}}},\ Vol.~\bibinfo {volume} {2}\ (\bibinfo
  {publisher} {CERN, European Organization for Nuclear Research},\ \bibinfo
  {year} {1964})\BibitemShut {NoStop}%
\bibitem [{\citenamefont {Møller}\ and\ \citenamefont
  {Schmiedmayer}(2020)}]{10.21468/SciPostPhys.8.3.041}%
  \BibitemOpen
  \bibfield  {author} {\bibinfo {author} {\bibfnamefont {F.~S.}\ \bibnamefont
  {Møller}}\ and\ \bibinfo {author} {\bibfnamefont {J.}~\bibnamefont
  {Schmiedmayer}},\ }\bibfield  {title} {\bibinfo {title} {{Introducing
  i{F}luid: a numerical framework for solving hydrodynamical equations in
  integrable models}},\ }\href {https://doi.org/10.21468/SciPostPhys.8.3.041}
  {\bibfield  {journal} {\bibinfo  {journal} {SciPost Phys.}\ }\textbf
  {\bibinfo {volume} {8}},\ \bibinfo {pages} {41} (\bibinfo {year}
  {2020})}\BibitemShut {NoStop}%
\bibitem [{\citenamefont {Korepin}\ \emph {et~al.}(1993)\citenamefont
  {Korepin}, \citenamefont {Bogoliubov},\ and\ \citenamefont
  {Izergin}}]{korepin_bogoliubov_izergin_1993}%
  \BibitemOpen
  \bibfield  {author} {\bibinfo {author} {\bibfnamefont {V.~E.}\ \bibnamefont
  {Korepin}}, \bibinfo {author} {\bibfnamefont {N.~M.}\ \bibnamefont
  {Bogoliubov}},\ and\ \bibinfo {author} {\bibfnamefont {A.~G.}\ \bibnamefont
  {Izergin}},\ }\href {https://doi.org/10.1017/CBO9780511628832} {\emph
  {\bibinfo {title} {Quantum Inverse Scattering Method and Correlation
  Functions}}},\ Cambridge Monographs on Mathematical Physics\ (\bibinfo
  {publisher} {Cambridge University Press},\ \bibinfo {year}
  {1993})\BibitemShut {NoStop}%
\end{thebibliography}%
